\documentclass[journal,12pt,onecolumn]{IEEEtran}
\ifCLASSINFOpdf
\else
\fi
\usepackage{setspace}
\ifCLASSOPTIONonecolumn \doublespace \fi
\usepackage{amsfonts}
\usepackage{amsmath}
\usepackage{graphicx}
\usepackage{cite}
\usepackage{caption2}
\graphicspath{{Figures/}} \DeclareGraphicsExtensions{.eps,.ps}
\allowdisplaybreaks[1]

\begin{document}
%
\title{Power Allocation for Precoding in Large-Scale MIMO Systems with Per-Antenna Constraint}

\author{Dengkui~Zhu,~Boyu~Li,~\IEEEmembership{Member,~IEEE,}~and~Ping~Liang
\thanks{
The authors are with RF DSP Inc., 30 Corporate Park, Suite 210, Irvine, CA 92606, USA (e-mail: dkzhu@rfdsp.com; byli@rfdsp.com; pliang@rfdsp.com).

P. Liang is also with University of California - Riverside, Riverside, CA 92521, USA (e-mail: liang@ee.ucr.edu). 
}}


%


\maketitle
   
\begin{abstract}
Large-scale MIMO systems have been considered as one of the possible candidates for the next-generation wireless communication technique, due to their potential to provide significant higher throughput than conventional wireless systems. For such systems, Zero-Forcing (ZF) and Conjugate Beamforming (CB) precoding have been considered as two possible practical spatial multiplexing techniques, and their average achievable sum rates have been derived on the sum power constraint. However, in practice, the transmitting power at a base station is constrained under each antenna. In this case, the optimal power allocation is a very difficult problem. In this paper, the suboptimal power allocation methods for both ZF-based and CB-based precoding in large-scale MIMO systems under per-antenna constraint are investigated, which could provide useful references for practice.   
\end{abstract}


%

\section{Introduction}

Large-scale Multiple-Input Multiple-Output (MIMO) systems \cite{Marzetta_massive_MIMO_original,Hoydis_LPComp,Hoydis_massive_MIMO,Rusek_massive_MIMO_overview,Bjornson_arXiv,Larsson_massive_MIMO_overview} have drawn substantial interests by both academia and industry
. 
In such systems, each Base Station (BS) is equipped with dozens to several hundreds transmitting antennas. One main advantage of large-scale MIMO systems is the potential capability to offer linear capacity growth without increasing power or bandwidth  
by employing Multi-User MIMO (MU-MIMO) to achieve the significant higher spatial multiplexing gains than conventional systems \cite{Marzetta_massive_MIMO_original,Hoydis_LPComp,Hoydis_massive_MIMO,Rusek_massive_MIMO_overview,Bjornson_arXiv,Larsson_massive_MIMO_overview}. In such systems, a BS selects multiple User-Equipments (UEs) at each scheduling slot and transmits data to them on the same time-frequency resource. 

The downlink of a MU-MIMO system can be modeled as the MIMO broadcast channel \cite{Yu_PAPC}. With a nonlinear precoding technique Dirty Paper Coding (DPC) \cite{Costa_DPC}, under the Sum Power Constraint (SPC), the sum capacity of the MIMO broadcast channel can be achieved \cite{Caire_MIMO_Broadcast,Viswanath_Broadcast,Yu_Broadcast}. However, the nonlinear DPC requires substantial complexity at both transmitter and receiver \cite{Lee_DPC_LP}, which is difficult to implement in practice
. In addition, for a practical BS, each antenna has its own power amplifier, which means that the power allocation is under the Per-Antenna Power Constraint (PAPC) instead of the SPC. In this case, the optimal power allocation and precoding method to maximize the sum rate is an even more difficult problem. Recently, in \cite{Christopoulos_PAPC}, a solution was provided for this issue with the complexity of $O(M^6)$ where $M$ is the number of BS antennas, which is obviously unrealistic for large-scale MIMO systems equipped with many antennas, e.g., $M=128$. As a result, the motivation of this paper is to seek suboptimal solutions under the PAPC with both practical complexity and acceptable sum rates for large-scale MIMO systems. Note that the solution provided in \cite{Yu_PAPC} minimizes the per-antenna power on each transmitting antenna
while enforces a set of Signal-to-Interference-plus-Noise Ratio (SINR) constraints on each UE, which is a different optimization problem from this paper.  

For large-scale MIMO systems, where $M$ is much larger than the number of UEs $K$ \cite{Hoydis_LPComp,Hoydis_massive_MIMO,Bjornson_arXiv,Rusek_massive_MIMO_overview}, it has been shown that under the SPC, the Zero-Forcing (ZF) precoding \cite{Wiesel_ZFP,Shepard_Argos,Hoydis_LPComp,Hoydis_massive_MIMO,Bjornson_arXiv,Rusek_massive_MIMO_overview,Marzetta_ZF_CB}, which removes the interference among the grouped UEs, can achieve the sum rate very close to the capacity-achieving DPC, hence it is virtually optimal \cite{Rusek_massive_MIMO_overview}. 
Note that as a linear precoding technique, ZF precoding requires significantly less complexity than the nonlinear DPC. Therefore, it has been considered as one of the potential practical precoding methods for large-scale MIMO systems \cite{Shepard_Argos,Hoydis_LPComp,Hoydis_massive_MIMO,Bjornson_arXiv,Rusek_massive_MIMO_overview,Marzetta_ZF_CB}.
As a result, instead of seeking the optimal power allocation and precoding method to maximize the sum rate under the PAPC for large-scale MIMO systems, an alternative strategy is to seek suboptimal power allocation methods based on ZF precoding that can achieve sum rates close to ZF precoding under the SPC with practical complexity. 
In \cite{Wiesel_ZFP}, the throughput optimization problem of ZF precoding under the PAPC for conventional MU-MIMO systems was modeled as a standard Determinant Maximization (MAXDET) program subject to linear matrix inequalities \cite{Vandenberghe_MAXDET}, which has the complexity of $O(M^{4.5}K^2+M^{2.5}K^4)$. Similarly to the solution provided in \cite{Christopoulos_PAPC}, the complexity is too high for large-scale MIMO systems where $M$ is very large in practice. In \cite{Boccardi_ZF_PAPC,Ohwatari_PAPC}, to optimize the throughput of ZF precoding under the PAPC for conventional MU-MIMO systems, a water-filling based method was applied, whose complexity and performance will be compared with our proposed methods for large-scale MIMO systems. 

Conjugate Beamforming (CB) precoding \cite{Marzetta_massive_MIMO_original,Hoydis_LPComp,Hoydis_massive_MIMO,Bjornson_arXiv,Rusek_massive_MIMO_overview,Shepard_Argos,Marzetta_ZF_CB} is another potential practical precoding method for large-scale MU-MIMO systems due to its simplicity for implementation. The average achievable sum rate of CB under the SPC has been derived in \cite{Marzetta_massive_MIMO_original,Hoydis_LPComp,Hoydis_massive_MIMO,Rusek_massive_MIMO_overview,Marzetta_ZF_CB}. Similarly to ZF, however, CB also faces the power allocation problem under the PAPC.

In this paper, we address the power allocation problem of ZF-based precoding under the PAPC according to two criteria, i.e., maximum power utilization and minimum multi-user interference, in order to achieve the performance close to ZF under the SPC with practical complexity in large-scale MIMO systems. Specifically, based on the first criterion, an orthogonal projection method with the complexity of $O(MK)$ and a feasible Newton iterative method with the complexity of $O(m_{\mathrm{FNI}}MK)$, where $m_{\mathrm{FNI}}$ is the number of iterations, are presented. For the second criterion, a linear scaling method with the complexity of $O(MK)$ and a Newton iterative method with the complexity of $O[m_{\mathrm{NI}}(MK^2+K^3)]$, where $m_{\mathrm{NI}}$ is the number of iterations, are presented. Simulation results show that with relatively accurate Channel State Information (CSI), i.e., the correlation coefficient between ideal and measured CSI $\beta$ is no less than $0.9$, the first criterion performs better at the relatively low Signal-to-Noise Ratio (SNR) region, and the feasible Newton iterative method achieves the performance close to ZF under the SPC, and significantly better than the orthogonal projection method with an acceptable increase in complexity. On the contrary, the second criterion is the better choice at the relatively high SNR region, and the linear scaling method with very low complexity achieves the performance close to the high complexity Newton iterative method and the water-filling based method employed in \cite{Boccardi_ZF_PAPC,Ohwatari_PAPC}, and with only a small loss compared to ZF under the SPC. In the case of relatively inaccurate CSI, i.e., $\beta < 0.9$, the feasible Newton iterative method provides the best performance, which is close to ZF under the SPC, with acceptable complexity regardless of the SNR. In addition, an accurate performance approximation to the linear scaling method is provided. Furthermore, we provide a simple power allocation method for CB-based precoding under the PAPC and prove that this method suffers little throughput loss compared to CB under the SPC. The results of this paper could provide useful references for practical large-scale MIMO systems. 

The remainder of this paper is organized as follows. Section II briefly describes the system model. In Section III and Section IV, the power allocation problem of ZF-based precoding under the PAPC is addressed according to the maximum power utilization and minimum multi-user interference criteria respectively. Then, Section V discusses the computation complexity of the proposed algorithms presented in the previous two sections. In Section VI, a simple power allocation method for CB-based precoding under the PAPC is presented and proved with little throughput loss compared to CB under the SPC. In Section VII, simulation results are provided. Finally, a conclusion is drawn in Section VIII.



 
\section{System Model}
Consider a large-scale MU-MIMO wireless system, where the BS with $M$ transmitting antennas serves $K$ UEs with single receiving antenna on each radio resource, e.g., a subcarrier or a OFDM symbol. Let $\mathbf{h}_k$ be the $M \times 1$ channel vector of the $k$th UE, then the $K\times M$ effective channel matrix is $\mathbf{H} = \left[\mathbf{h}_1 \, \mathbf{h}_2 \, \cdots \, \mathbf{h}_K \right]^{\mathrm{T}} $. The channel is assumed to be in uncorrelated Rayleigh fading, i.e., the elements of $\mathbf{H}$ are independent and identically distributed (i.i.d.) zero-mean unit-variance complex Gaussian variables, which is the same assumption as \cite{Marzetta_massive_MIMO_original,Hoydis_LPComp,Hoydis_massive_MIMO,Rusek_massive_MIMO_overview,Bjornson_arXiv,Marzetta_ZF_CB} for analyzing large-scale MIMO systems. 
The $M\times K$ precoding matrix employed by the BS 
is represented by $\mathbf{W} = \left[\mathbf{w}_1 \, \cdots \,\mathbf{w}_K \right]$, where ${\mathbf{w}}_k \in \mathbb{C}^{M \times 1}$, $k=1,\cdots,K$. The precoding matrix  $\mathbf{W}$ can be rewritten as $\mathbf{w}_k = \sqrt{\mathbf{p}_k} \circ e^{j\boldsymbol{\theta}_k}$, where $\mathbf{p}_k$ is the power allocation vector with $p_{mk}$ being the power allocated for the $k$th UE on the $m$th antenna, $\boldsymbol{\theta}_k$ is the phase vector for the $k$th UE, and $\circ$ denotes the Hadamard product \cite{Horn_Matrix_Analysis}. Considering the PAPC instead of the SPC because of the independent power amplifier in practical systems, to seek the optimal precoding matrix $\mathbf{W}$ under the PAPC is equivalent to the optimal problem below as
\begin{equation}
\label{f52} 
\begin{split}
&\mathop {\max }\limits_{\left\{ \mathbf{w}_k \right\}_{k = 1}^K} \sum\limits_{k = 1}^K {\log }_2 \left( 1 + \gamma _k \right), \\
&\mathrm{s.t: \,\,} \gamma _k = \frac{\left| \mathbf{h}^{\mathrm{T}}_k \mathbf{w}_k \right|^2}{\sum\nolimits_{j = 1,j \ne k}^K \left| \mathbf{h}^{\mathrm{T}}_k \mathbf{w}_j 
\right|^2 + \sigma _{n,k}^2 },\\
&\mathrm{and \,\,} \left[ \mathbf{W}\mathbf{W}^{\mathrm{H}} \right]_{m.m}\le P_{\mathrm{Ant}, m}, \,\, m= 1, \cdots ,M,
\end{split}
\end{equation}
where $\sigma_{n,k}^2$ denotes the noise variance for the $k$th UE, $[ \mathbf{W}\mathbf{W}^{\mathrm{H}} ]_{m.m}$ denotes the $m$th diagonal element of the $M \times M$ matrix product $\mathbf{W}\mathbf{W}^{\mathrm{H}}$, and $P_{\mathrm{Ant}, m}$ is the maximal power for the $m$th antenna. Unfortunately, Problem (\ref{f52}) is not convex and only approximated solutions can be obtained through very complicated iterative search, e.g., the complexity of the solution proposed in \cite{Christopoulos_PAPC} is about $O(M^6)$, which is unrealistic for practical systems, especially in large-scale MIMO systems where $M$ is very large, e.g., more than $100$. Hence, we provide suboptimal solutions with practical complexity to this problem by exploiting the special properties of large-scale MIMO systems.
 
\subsection{ZF-Based Power Allocation}
Note that the channel capacity under the SPC is no less than the PAPC as the feasible domain of the latter problem is a subset of the former problem. Moreover, the simple linear ZF precoding method could achieve the throughput very close to the multi-user channel capacity under the SPC considering that $M/K$ is very large for large-scale MIMO systems, e.g., more than $10$ \cite{Rusek_massive_MIMO_overview}. Its precoding matrix is given by
\begin{equation}
\label{f53}
\mathbf{W}^{\mathrm{ZF-SPC}} = \mathbf{H}^\mathrm{H}\left( \mathbf{H} \mathbf{H}^{\mathrm{H}} \right)^{ - 1} \sqrt{\mathbf{\Phi}^{\mathrm{ZF-SPC}}} = {\mathbf{\Xi}}^{\mathrm{ZF-SPC}} \circ e^{j\mathbf{\Theta}^{\mathrm{ZF-SPC}}}, 
\end{equation}
where $\mathbf{\Phi}^{\mathrm{ZF-SPC}}$ is a diagonal matrix with $\phi^{\mathrm{ZF-SPC}} _k$ being the scaling factor of the $k$th UE to satisfy the SPC, $\mathbf{\Xi}^\mathrm{ZF-SPC}=[\sqrt{p_{mk}^{\mathrm{ZF-SPC}}}]$ is the square root matrix of the power allocation matrix $\mathbf{P}^\mathrm{ZF-SPC}$, and $\mathbf{\Theta}^{\mathrm{ZF-SPC}}$ is the phase matrix. Note that for large-scale MIMO systems, $\mathbf{\Phi}^{\mathrm{ZF-SPC}}$ is given by \cite{Rusek_massive_MIMO_overview} as $\mathbf{\Phi}^{\mathrm{ZF-SPC}} = \phi^{\mathrm{ZF-SPC}} \mathbf{I}_K$, where $\phi^{\mathrm{ZF-SPC}} = 1 / \mathrm{Tr} [( \mathbf{H} \mathbf{H}^{\mathrm{H}} )^{-1}]$ with the total maximal transmitting power being assumed to be $1$ in this paper, and $\mathbf{I}_K$ denotes the $K$-dimensional identity matrix. Hence, if we could find a precoding matrix under the PAPC which has the minimal Euclid distance from the ZF precoding matrix under the SPC, then we obtain a suboptimal solution under the PAPC. Let $\mathbf{W}^{\mathrm{ZF-PAPC}} = \mathbf{\Xi}^{\mathrm{ZF-PAPC}} \circ e^{j\mathbf{\Theta}^{\mathrm{ZF-PAPC}}}$ denote a feasible precoding matrix under the PAPC, then we have the following lemma.

\emph {Lemma 1:} Given the Frobenius norm of $\mathbf{W}^{\mathrm{ZF-PAPC}}$, the Frobenius norm of $\| \mathbf{W}^{\mathrm{ZF-SPC}} - \mathbf{W}^\mathrm{ZF-PAPC} \|_\mathrm{F}^2$ reaches the minimum value when $\mathbf{\Theta}^{\mathrm{ZF-PAPC}} = \mathbf{\Theta}^{\mathrm{ZF-SPC}}$.
\begin{IEEEproof} After expanding $\| \mathbf{W}^{\mathrm{ZF-SPC}} - \mathbf{W}^{\mathrm{ZF-PAPC}} \|_\mathrm{F}^2$, we have
\begin{align}
\label{f1} 
\left\| \mathbf{W}^{\mathrm{ZF-SPC}} - \mathbf{W}^{\mathrm{ZF-PAPC}} \right\|_\mathrm{F}^2 & = \left\| \mathbf{W}^{\mathrm{ZF-SPC}} \right\|_\mathrm{F}^2 + \left\| \mathbf{W}^{\mathrm{ZF-PAPC}} \right\|_\mathrm{F}^2 - 2 \Re \left[ \mathrm{Tr}\left( \mathbf{W}^{\mathrm{ZF-SPC,H}} \mathbf{W}^{\mathrm{ZF-PAPC}} \right) \right] \nonumber \\
& \ge \left\| \mathbf{\Xi}^{\mathrm{ZF-SPC}} - \mathbf{\Xi}^{\mathrm{ZF-PAPC}} \right\|_\mathrm{F}^2,
\end{align} 
where the equality holds only when $\mathbf{\Theta}^{\mathrm{ZF-PAPC}} = \mathbf{\Theta}^{\mathrm{ZF-SPC}}$ based on the Cauchy-Schwarz inequality \cite{Horn_Matrix_Analysis}. 
\end{IEEEproof}
Based on Lemma $\mathrm{1}$, the suboptimal precoding matrix under the PAPC can be written as $\mathbf{W}^{\mathrm{ZF-PAPC}} = \mathbf{\Xi}^{\mathrm{ZF-PAPC}} \circ e^{j\mathbf{\Theta}^{\mathrm{ZF-SPC}}}$ where $\mathbf{\Xi}^{\mathrm{ZF-PAPC}}$ is selected to minimize $d = \| \mathbf{\Xi}^{\mathrm{ZF-PAPC}} - \mathbf{\Xi}^{\mathrm{ZF-SPC}} \|_\mathrm{F}^2$. For the virtually optimal power allocation vector $\mathbf{P}^{\mathrm{ZF-SPC}}$, it ensures full power utilization while guarantees zero multi-user interference. For the PAPC, the minimum value of $d$ can be achieved if $\mathbf{P}^{\mathrm{ZF-PAPC}} = \mathbf{\Xi}^{\mathrm{ZF-PAPC}} \circ \mathbf{\Xi}^{\mathrm{ZF-PAPC}}$ satisfies the same conditions as $\mathbf{P}^{\mathrm{ZF-SPC}}$. Specifically, for full power utilization, $\mathbf{P}^{\mathrm{ZF-PAPC}}$ is constrained by 
\begin{equation}
\label{f2}
\sum\limits_{k = 1}^K p_{mk}^{\mathrm{ZF-PAPC}} = \frac{1}{M} , \,\, m = 1, \cdots ,M
\end{equation}
where each antenna is assumed to have the same maximal power of $1/M$. To achieve zero multi-user interference by multi-user beamforming, the ratio of the allocated power for the $K$ UEs on each antenna needs to be the same as $\mathbf{P}^{\mathrm{ZF-SPC}}$ to retain the orthogonality among UEs. Otherwise, the orthogonality among UEs will be violated, resulting multi-user interference. Hence, $\mathbf{P}^{\mathrm{ZF-PAPC}}$ is constrained by     
\begin{equation}
\label{f3}
\frac{p_{mk}^{\mathrm{ZF-PAPC}} }
{ p_{mk}^{\mathrm{ZF-SPC}} } = \frac{p_{lk}^{\mathrm{ZF-PAPC}}}
{p_{lk}^{\mathrm{ZF-SPC}} } = \frac{\sum_{i=1}^M p_{ik}^{\mathrm{ZF-PAPC}}}
{\sum_{i=1}^M p_{ik}^{\mathrm{ZF-SPC}} },\,\,m,l = 1, \cdots ,M,\,\,k=1,\cdots,K.
\end{equation}
Unfortunately, no explicit solution exists to satisfy (\ref{f2}) and (\ref{f3}) simultaneously as the number of equations is more than the unknown parameters. Hence, we relax the two constraints by dividing them into two classes, where one is the strong constraint that has to be satisfied while the other is the loose constraint that allows distortion. In fact, $\mathbf{P}^{\mathrm{ZF-PAPC}}$ approaches to $\mathbf{P}^{\mathrm{ZF-SPC}}$ in two different directions based on the two classifications, i.e., maximum power utilization and minimum multi-user interference. In sections III and IV, we find $\mathbf{P}^{\mathrm{ZF-PAPC}}$ based on these two different constraint classifications.

\subsection{CB-Based Power Allocation}
Even though CB is not a candidate solution to Problem (\ref{f52}), it is still considered as a potential precoding method for large-scale MIMO systems due to its simplicity. Similarly to ZF, it also faces the problem on how to choose the power allocation matrix $\mathbf{W}^{\mathrm{CB-PAPC}}$ under the PAPC so that the performance loss is minimized compared to the SPC. This issue will be discussed in Section VI.

\section{Maximum Power Utilization}
For the maximum power utilization criterion, (\ref{f2}) is selected as the strong constraint, i.e., seeking the best power allocation to relieve multi-user interference due to full power usage of each antenna. Equation (\ref{f2}) is rewritten in matrix form as
\begin{equation}
\label{f4}
\mathbf{A} \mathbf{x} = \mathbf{b}
\end{equation}
where
\ifCLASSOPTIONonecolumn
\begin{equation}
\label{f5}
\mathbf{A} = \left[ \begin{array}{*{20}c} \mathbf{I}_M & \cdots & \mathbf{I}_M 
\end{array} \right],
\end{equation}
\else
\begin{equation}
\renewcommand{\arraycolsep}{3.5pt}
\renewcommand{\arraystretch}{0.8}
\label{f5}
\mathbf{A} = \left[ \begin{array}{*{20}c} \mathbf{I}_M & \cdots & \mathbf{I}_M 
\end{array} \right],
\end{equation}
\fi
\begin{equation}
\label{f6}
\mathbf{b} = \left[ \begin{array}{*{20}c} \frac{1}{M} & \cdots & \frac{1}{M}\end{array}  \right]^{\mathrm{T}},
\end{equation}
and
\ifCLASSOPTIONonecolumn
\begin{align}
\label{f7}
\mathbf{x} = \mathrm{vec}\left( \mathbf{P}^{\mathrm{ZF-PAPC}} \right) = \left[ \begin{array}{*{20}{c}}
p_{11}^{\mathrm{ZF-PAPC}}& \cdots &p_{M1}^{\mathrm{ZF-PAPC}} & \cdots &p_{1K}^{\mathrm{ZF-PAPC}}& \cdots &p_{MK}^{\mathrm{ZF-PAPC}}
\end{array} \right]^{\mathrm{T}}. 
\end{align}
\else
\begin{equation}
\renewcommand{\arraycolsep}{1.5pt}
\label{f7}
\mathbf{x} = \mathrm{vec}\left( {\mathbf{P}}^{\mathrm{ZF-PAPC}} \right) = \left[ \begin{array}{*{20}{c}}
p_{11}^{\mathrm{ZF-PAPC}}& \cdots &p_{M1}^{\mathrm{ZF-PAPC}} & \cdots &p_{1K}^{\mathrm{ZF-PAPC}}& \cdots & p_{MK}^{\mathrm{ZF-PAPC}}
\end{array} \right]^{\mathrm{T}}.
\end{equation}
\fi
Note that $\mathbf{A}$ is a $M \times MK$ matrix, $\mathbf{b}$ is a $M \times 1$ vector,  and $\mathrm{vec} ( {\mathbf{P}}^{\mathrm{ZF-PAPC}})$ means stacking the column vectors of $\mathbf{P}^{\mathrm{ZF-PAPC}}$. Since $\mathbf{x}$ is the power allocation vector, it satisfies the condition ${\mathbf{x}}\succeq 0$. Hence, minimizing $d = \| \mathbf{\Xi}^{\mathrm{ZF-PAPC}} - \mathbf{\Xi}^{\mathrm{ZF-SPC}} \|_\mathrm{F}^2$ equals to
\begin{equation}
\label{f8}
\min \left\| \sqrt{\mathbf{x}} - \sqrt{\mathbf{r}} \right\|_2^2 
\end{equation}
where
\ifCLASSOPTIONonecolumn
\begin{equation}
\label{f9}
\mathbf{r} = \mathrm{vec}\left( \mathbf{P}^{\mathrm{ZF-SPC}} \right) = \left[ \begin{array}{*{20}{c}}
p_{11}^{\mathrm{ZF-SPC}}& \cdots &p_{M1}^{\mathrm{ZF-SPC}} & \cdots &p_{1K}^{\mathrm{ZF-SPC}}& \cdots &p_{MK}^{\mathrm{ZF-SPC}}
\end{array} \right]^{\mathrm{T}}.
\end{equation}
\else
\begin{equation}
\renewcommand{\arraycolsep}{2pt}
\label{f9}
\mathbf{r} = \mathrm{vec}\left( \mathbf{P}^{\mathrm{ZF-SPC}} \right) = \left[ \begin{array}{*{20}{c}}
p_{11}^{\mathrm{ZF-SPC}}& \cdots &p_{M1}^{\mathrm{ZF-SPC}} & \cdots &p_{1K}^{\mathrm{ZF-SPC}}& \cdots &p_{MK}^{\mathrm{ZF-SPC}}
\end{array} \right]^{\mathrm{T}}.
\end{equation}
\fi

In fact, (\ref{f8}) is a loose condition of (\ref{f3}), which minimizes 
the multi-user interference
. 
In summary, the maximum power utilization method equals to the optimization problem 
\begin{equation}
\label{f11}
\begin{split}
&\min \left\| \sqrt{\mathbf{x}} - \sqrt{\mathbf{r}} \right\|_2^2,   \\
&\mathrm{s.t. \,\,} \mathbf{Ax} - \mathbf{b} = 0, \\
&\qquad - \mathbf{x} \preceq  0. \\
\end{split}
\end{equation}
Next, we provide two ways to solve this problem in the following subsections A and C respectively.

\subsection{Orthogonal Projection Method}
The first method is to obtain a suboptimal solution based on orthogonally projecting the vector $\mathbf{r}$ into the real affine subspace of $\mathbf{x}$, where $\| \sqrt \mathbf{x} - \sqrt {\mathbf{r}}  \|_2^2$ is approximated by $\| \mathbf{x - r} \|_2^2$ in (\ref{f11}). Note that $\mathbf{A}$ has full row rank, hence the solution of (\ref{f4}) is in a subspace with the dimension of $MK-M$. As a result, the subspace of solutions to the equation
$\mathbf{Ax} - \mathbf{b} = \mathbf{0}$ is constructed by rewriting it in an augmented matrix form as
\begin{equation}
\renewcommand{\arraystretch}{0.8}
\label{f12}
\left[ {\begin{array}{*{20}c}
   {\mathbf{A}} & {\mathbf{b}}  \\
 \end{array} } \right]\left[ {\begin{array}{*{20}c}
   {\mathbf{x}}  \\
   { - 1}  \\
 \end{array} } \right] = 0.
\end{equation}
Obviously, all vectors with the form of $ \tilde{\mathbf{x}} = \rho [\mathbf{x}, \,\, -1]^{\mathrm{T}} $, where $\rho$ is a real-valued scaling factor, constitute the null subspace of matrix 
$\tilde{\mathbf{A}} = [ \mathbf{A} \,\, \mathbf{b} ]$. Therefore, a vector $ \tilde{\mathbf{r}} = [ \mathbf{r}^{\mathrm{T}}, \,\,  - a ]^{\mathrm{T}} $ can be projected into this null subspace by
\begin{equation}
\label{f13}
\tilde{\mathbf{p}}^{\mathrm{Sub}} = \left( \mathbf{I}_{MK+1} - \tilde{\mathbf{A}}^{\mathrm{T}} \left( \tilde{\mathbf{A}} \tilde{\mathbf{A}}^{\mathrm{T}} \right)^{ - 1}\tilde{\mathbf{A}} \right) \tilde{\mathbf{r}}.
\end{equation}
Then, the elements of $\tilde{\mathbf{p}}^{\mathrm{Sub}} $ are normalized by the negative of the last element and are denoted by a vector $\mathbf{p}^{\mathrm{Sub}}$ as
\begin{equation}
\label{f14}
\mathbf{p}^{\mathrm{Sub}} \left( k \right) = \frac{\tilde{\mathbf{p}}^{\mathrm{Sub}} \left( k \right)}
{-\tilde{\mathbf{p}}^{\mathrm{Sub}} \left( MK + 1 \right)}, \,\,k = 1, \cdots ,KM+1.
\end{equation}
Hence, ${\mathbf{p}}^{\mathrm{Sub}} \left( k \right)$ has a form of $ [ \mathbf{x}, \,\, -1 ]^{\mathrm{T}} $ and it is a solution of (\ref{f12}).

\subsection{Analysis on the Value of $a$}
The parameter $a>0$ in the vector $\tilde{\mathbf{r}}$ ensures that the final solution in (\ref{f14}) satisfies $\mathbf{p}^{\mathrm{Sub}} \succeq 0$. In this subsection, we provide the analyses on how to choose a reasonable value of $a$. Let $\mathbf{G}_{\tilde A} = \tilde{\mathbf{A}} \tilde{\mathbf{A}}^{\mathrm{T}}$, then $\mathbf{G}_{\tilde A}$ can be written into
\begin{equation}
\label{f54}
{{\mathbf{G}}_{\tilde A}} = {\mathbf{A}}{{\mathbf{A}}^{\mathrm{T}}} + {\mathbf{b}}{{\mathbf{b}}^{\mathrm{T}}} = K{{\mathbf{I}}_M} + \frac{1}{M^2}\mathbf{E}_M
\end{equation}
where $\mathbf{E}_M$ denotes a $M \times M$ matrix with all elements being $1$. Since $M$ is very large, e.g., more than $100$, the inverse matrix of $\mathbf{G}_{\tilde A}$ can be estimated as
\begin{equation}
\label{f55}
\mathbf{G}_{\tilde A}^{ - 1} \approx \frac{1}{K} \sum_{n=0}^{\infty} \left(\mathbf{I}_M - \frac{1}{K} \mathbf{G}_{\tilde A} \right)^n = \frac{1}{K} \sum_{n=0}^{\infty} \left( - \frac{1}{M^2K} \mathbf{E}_M \right)^n \approx \frac{1}{K}{\mathbf{I}_M} - \frac{1}{M^2K^2}\mathbf{E}_M
\end{equation}
with negligible error according to the Neumann series expansion of the inverse matrix \cite{NeumannSeries}, e.g., less than $1/M^2K^2$ per element. With (\ref{f55}), after block matrix multiplications, the projection matrix $\mathbf{M}^{\mathrm{Proj}} = \mathbf{I}_{MK+1} - \tilde{\mathbf{A}}^{\mathrm{T}} (\tilde{\mathbf{A}} \tilde{\mathbf{A}}^{\mathrm{T}} )^{ - 1} \tilde{\mathbf{A}}$ becomes to
\begin{equation}
\label{f56}
\mathbf{M}^{\mathrm{Proj}} = \left[ \begin{array}{*{20}{c}}
\mathbf{D}_1&\mathbf{D}_2& \cdots &\mathbf{D}_2&\mathbf{v}\\
\mathbf{D}_2&\mathbf{D}_1& \cdots &\mathbf{D}_2&\mathbf{v}\\
 \vdots & \vdots & \ddots & \vdots & \vdots \\
\mathbf{D}_2&\mathbf{D}_2& \cdots &\mathbf{D}_1&\mathbf{v}\\
\mathbf{v}^{\mathrm{T}}&\mathbf{v}^{\mathrm{T}}& \cdots & \mathbf{v}^{\mathrm{T}}& c
\end{array} \right],
\end{equation}
where 
\begin{equation}
\label{f71}
\mathbf{D}_1 = \left( 1 - \frac{1}{K} \right) \mathbf{I}_M + \frac{1} {M^2K^2}\mathbf{E}_M,
\end{equation}
\begin{equation}
\label{f57}
\mathbf{D}_2 =  - \frac{1}{K} \mathbf{I}_M + \frac{1}{M^2K^2} \mathbf{E}_M,
\end{equation}
\begin{equation}
\label{f68}
\mathbf{v} = \left[ {\begin{array}{*{20}{c}}
\frac{1}{M^2K^2} - \frac{1}{MK} & \cdots &\frac{1}{M^2K^2} - \frac{1}{MK}
\end{array}} \right]^{\mathrm{T}},
\end{equation}
and
\begin{equation}
\label{f72}
c = 1 + \frac{1}{MK} - \frac{1}{M^2K^2}.
\end{equation}
Substituting (\ref{f56})-(\ref{f72}) into (\ref{f13}), the $k^{th}$ element of $\tilde{\mathbf{p}}^{\mathrm{Sub}}$ is given by
\begin{equation}
\label{f58}
\tilde{\mathbf{p}}^{\mathrm{Sub}} \left( k \right) = \left( 1 - \frac{1}{K} \right)p_k^{\mathrm{ZF-SPC}} - \frac{1}{K}\sum\limits_{l \in \Omega _{k,1}} p_l^{\mathrm{ZF-SPC}} + \frac{1}{M^2K^2} \sum\limits_{l=1}^{MK} p_l^{\mathrm{ZF-SPC}} + a\left( \frac{1}{MK} - \frac{1}{M^2K^2}\right),
\end{equation}
with $k=1,\cdots,KM$, where $\Omega _{k,1} = \{ k + M, \cdots ,k + ( K - 1)M \}_{\bmod MK}$. 
Since the terms $p_l^{\mathrm{ZF-SPC}}$, $l=1,\cdots,KM$, are non-negative numbers and $\sum_{l \in \Omega _{k,1}} p_l^{\mathrm{ZF-SPC}}  \le 1$, the relation $\tilde{\mathbf{p}}^{\mathrm{Sub}} \succeq 0$ would be guaranteed when $a \ge (M^2K)/(MK-1) $. Hence, $a$ can be chosen as any number no less than $(M^2K)/(MK-1)$.

Note that on the one hand, $\| \sqrt {\mathbf{x}} - \sqrt{\mathbf{r}} \|_2^2$ is not necessarily equivalent to $\| \mathbf{x - r} \|_2^2$. On the other hand, $\tilde{\mathbf{p}}^{\mathrm{Sub}}$ is a solution to (\ref{f12}) but it does not necessarily have the form of $ [ \mathbf{x}, \,\, -1 ]^{\mathrm{T}} $. Therefore, $\mathbf{p}^{\mathrm{Sub}}$ is only a suboptimal solution to the original problem (\ref{f11}).

\subsection{Feasible Newton Iterative Method}
In this subsection, we solve the optimization problem (\ref{f11}) by iterative search. Since the objective function is convex and the constraint is an affine function, a globally optimal point exists. With the interior-point method \cite{Boyd_CO}, (\ref{f11}) is transformed into the equivalent problem by introducing a parameter $t$ as
\begin{equation}
\label{f15}
\begin{split}
& \text{min}f\left( t,\mathbf{x} \right) =  - t\left\| \sqrt{\mathbf{x}} -\sqrt{ \mathbf{r}} \right\|_2^2  - \sum\limits_{i = 1}^{MK} \log x_i,  \\
& \mathrm{s.t. \,\,} \mathbf{Ax = b}. 
\end{split}
\end{equation}
Given the value of $t$, (\ref{f15}) could be solved by feasible Newton iteration \cite{Boyd_CO}
. The gradient vector and Hessian matrix of $f( t,\mathbf{x} )$ are
\begin{equation}
\label{f16}
\nabla f_{\mathbf{x}} = t - \frac{t\sqrt {\mathbf{r}} }{\sqrt {\mathbf{x}} } - \frac{1}{\mathbf{x}}
\end{equation}
and
\begin{equation}
\label{f17}
\nabla ^2 f_{\mathbf{x}} = \left[ \begin{array}{*{20}{c}}
t\sqrt{r_1}x_1^{-\frac{3}{2}} + x_1^{-2} & & \\
 & \ddots & \\
 & &t\sqrt{r_{MK}}x_{MK}^{-\frac{3}{2}} + x_{MK}^{- 2}
\end{array} \right]
\end{equation}
respectively. Then, the iteration process can be carried out as in \cite{Boyd_CO} 
. 
Note that the initial value could be chosen as (\ref{f14}) since it is already a suboptimal solution of the original problem. The initial value of $t$ should be chosen carefully to ensure the convergence. 
If the total maximal transmitting power is normalized to $1$, we could set it as $ MK\log _2 ( MK )$. Unfortunately, this value causes another problem, i.e., the matrix \renewcommand{\arraystretch}{0.7} \renewcommand{\arraycolsep}{2pt} $\left[ \begin{array}{*{20}c} \nabla ^2 f_{\mathbf{x}}  & \mathbf{A}^{\mathrm{T}}  \\\mathbf{A} & 0  \\
\end{array}  \right]$ employed in the iteration process becomes close to singular if $MK$ is large because the elements of $\mathbf A$ are either $0$ or $1$, which is very small relative to the elements of $\nabla ^2 f_{\mathbf{x}}$. This situation can be avoided by replacing the equation ${\mathbf{Ax = b}}$ with $MK \times {\mathbf{Ax}} = MK \times {\mathbf{b}}$ in the iteration process. To reduce the iteration number, setting $t = 1 / (MK \max_{\mathbf{x}} |\nabla f( x_i )|)$ 
is a good choice because it guarantees that the updated $\mathbf x$ would not distort the constraints while also ensures that the iteration converges rapidly.

\section{Minimum Multi-User Interference}
In contrast to Section III, in this section, (\ref{f3}) is selected as the strong constraint
, which means seeking the maximum power usage satisfying the requirement of no mutual interference among the grouped UEs.


Let $\mathbf{x}^{\mathrm{MMI}} = [x_1^{\mathrm{MMI}} \, \cdots x_K^{\mathrm{MMI}} \, ]^{\mathrm{T}} $ be the power allocation vector, where $x_k^{\mathrm{MMI}}$ is the total power allocated to the $k$th UE from summing all antennas. Then, according to (\ref{f3}), the power allocated to the $k$th UE on the $m$th antenna is 
\begin{equation}
\label{f59}
p_{mk}^{\mathrm{ZF-PAPC}} = \frac{p_{mk}^{\mathrm{ZF-SPC}}}{\alpha_k^{\mathrm{ZF-SPC}}} x_k^{\mathrm{MMI}} = a^{\mathrm{MMI}}_{mk} x_k^{\mathrm{MMI}}
\end{equation}
where $\alpha_k^{\mathrm{ZF-SPC}} = \sum\nolimits_{m = 1}^M p_{mk}^{\mathrm{ZF-SPC}} $ and $a^{\mathrm{MMI}}_{mk} = p_{mk}^{\mathrm{ZF-SPC}}/\alpha_k^{\mathrm{ZF-SPC}}$. With the strong constraint (\ref{f59}), 
the power usage can be written in the matrix form as
\begin{equation}
\label{f18}
\mathbf{A}^{\mathrm{MMI}}\mathbf{x}^{\mathrm{MMI}} - \mathbf{b} \preceq 0
\end{equation}
where the matrix $\mathbf{A}^{\mathrm{MMI}}$ is defined as
\ifCLASSOPTIONonecolumn
\begin{equation}
\label{f19}
\mathbf{A}^{\mathrm{MMI}} = \left[ \begin{array}{*{20}{c}}
a_{11}^{\mathrm{MMI}}& \cdots &a_{1K}^{\mathrm{MMI}}\\
 \vdots & \ddots & \vdots \\
a_{M1}^{\mathrm{MMI}}& \cdots &a_{MK}^{\mathrm{MMI}}
\end{array} \right]
\end{equation}
\else
\begin{equation}
\label{f19}
\mathbf{A}^{\mathrm{MMI}} = \left[ {begin{array}{*{20}{c}}
a_{11}^{\mathrm{MMI}}& \cdots &a_{1K}^{\mathrm{MMI}}\\
 \vdots & \ddots & \vdots \\
a_{M1}^{\mathrm{MMI}}& \cdots &a_{MK}^{\mathrm{MMI}}
\end{array} \right]
\end{equation}
\fi
 and $\mathbf{b}$ is the same as (\ref{f6}).
Therefore, minimizing the distance $d = \| \mathbf{\Xi}^{\mathrm{ZF-PAPC}} - \mathbf{\Xi}^{\mathrm{ZF-SPC}} \|_{\mathrm{F}}^2$ equals to
\begin{equation}
\label{f21}
\begin{split}
&\min \left\| \sqrt {\mathbf{p}^{\mathrm{MMI}}} - \sqrt{\mathbf{r}} \right\|_2^2,\\
&\mathrm{s.t.\,\,} \mathbf{p}^{\mathrm{MMI}} = \mathrm{vec}\left[ \mathbf{A}^{\mathrm{MMI}} \mathrm{diag} \left( \mathbf{x}^{\mathrm{MMI}} \right) \right],\\
&\quad \,\, \, \mathbf{A}^{\mathrm{MMI}} \mathbf{x}^{\mathrm{MMI}} - \mathbf{b} \preceq 0,\\
&\quad \,\, - \mathbf{x}^{\mathrm{MMI}} \preceq 0,
\end{split}
\end{equation}
where $\mathrm{diag} ( \mathbf{x}^{\mathrm{MMI}})$ represents a diagonal matrix with the diagonal elements from the vector $\mathbf{x}^{\mathrm{MMI}}$ and $\mathbf{r}$ is the same as (\ref{f9}).

Similarly to Section III, we provide two solutions to (\ref{f21}) in the next two subsections respectively, where the first one is simple but suboptimal while the second is globally optimal but iterative.

Note that (\ref{f21}) aims to minimize the distance between the two precoding matrices under the PAPC and the SPC, where the distance is determined by the power difference allocated to each UE. This approach is based on the fact that the ZF precoding under the SPC achieves the performance close to the channel capacity for large-scale MIMO systems. Another approach is to directly try to optimize the achievable sum-rate only with the constraint (\ref{f3}) as
\begin{equation}
\label{f66}
\begin{split}
&\mathop {\max }\limits \sum\limits_{k = 1}^K {\log }_2 \left( 1 + \gamma _k \right), \\
&\mathrm{s.t: \,\,} \gamma _k = \frac{\left| \mathbf{h}^{\mathrm{T}}_k \mathbf{w}^{\mathrm{MMI}}_k \right|^2}{\sigma _{n,k}^2 }, \\
& \quad \quad \mathbf{A}^{\mathrm{MMI}} \mathbf{x}^{\mathrm{MMI}}- \mathbf{b}\preceq 0, \\
& \quad \quad - \mathbf{x}^{\mathrm{MMI}} \preceq 0.
\end{split}
\end{equation}
The analytic solution to (\ref{f66}) is a water-filling based optimization problem, which has been derived in \cite{Boccardi_ZF_PAPC,Ohwatari_PAPC}, and its performance and complexity will be compared to the approaches proposed in this paper.


\subsection{Linear Scaling Method}
In this subsection, we solve (\ref{f21}) by a simple linear scaling method. 
The power of each UE is initially allocated to each antenna according to $p^{\mathrm{ZF-SPC}}_{mk}$.
Then, the power of each antenna is scaled to ensure that the sum power on each antenna dose not exceed the allowable value $1/M$. The linear scaling method is summarized in Table I. As a result, the total power allocated to the $k^{th}$ UE from summing all antennas is 
\begin{equation}
\label{f60}
x_k^{\mathrm{MMI-LS}} = \frac{\alpha^{\mathrm{ZF-SPC}}_k}{M} \frac{1}{\max_{m = 1}^M \sum\nolimits_{k = 1}^K p^{\mathrm{ZF-SPC}}_{mk} }.
\end{equation}
Since the elements of $\mathbf{H}$ are i.i.d. zero-mean unit-variance complex Gaussian variables and $M$ is much larger than $K$, the elements of its pseudo inverse $\mathbf{H}^\dag$ still can be approximated as i.i.d. zero-mean random variables. As mentioned in Section II, $\boldsymbol{\Phi}^{\mathrm{ZF-SPC}} = \phi^{\mathrm{ZF-SPC}} \mathbf{I}_K$ in (\ref{f53}) where $\phi^{\mathrm{ZF-SPC}} = 1 / \mathrm{Tr} [( \mathbf{H} \mathbf{H}^{\mathrm{H}} )^{-1}]$ in large-scale MIMO systems. Hence, based on (\ref{f53}), the squared 2-norm of the $\{m,k\}$th element of $\mathbf{H}^\dag$ is $p^{\mathrm{ZF-SPC}}_{mk}/\phi$ for large-scale MIMO systems.
Then, the power ratio of the $k^{th}$ UE under the PAPC to the SPC is approximated as 
\begin{equation}
\label{f62}
\frac{x_k^{\mathrm{MMI-LS}}}{\alpha^{\mathrm{ZF-SPC}}_k} = \frac{1}{M \max_{m = 1}^{M}p_{mk}^{\mathrm{ZF-SPC} } } = \frac{\frac{1}{\phi^{\mathrm{ZF-SPC}}}}{M \max_{m = 1}^{M} \sum\nolimits_{k = 1}^K \frac{1}{\phi^{\mathrm{ZF-SPC}}} p_{mk}^{\mathrm{ZF-SPC}} } = \frac{\left\| \mathbf{H}^\dag \right\|_{\mathrm{F}}^2}{M Q_{\max }},
\end{equation}
where $Q_{\mathrm{max}}$ denotes the maximum squared $2$-norm of the row vectors of $\mathbf{H}^\dag$ and  $\| \mathbf{H}^\dag \|_{\mathrm{F}}^2 = \mathrm{Tr} [( \mathbf{H} \mathbf{H}^{\mathrm{H}} )^{-1}]$. Relation (\ref{f62}) indicates that the power 
ratio of the PAPC to the SPC can be approximated to be directly related to $\mathbf{H}^\dag$. 

\begin{table}[!t]
\caption{Linear Scaling Method of (\ref{f21}).}
\centering  
\renewcommand{\arraystretch}{1.8}
\begin{tabular}{lccc}  
\hline
Step 1: $\tilde{p}_{mk}^{\mathrm{MMI-LS}} = p_{mk}^{\mathrm{ZF-SPC}}, \,\, m = 1, \cdots ,M, \,\, k = 1, \cdots ,K.$ \\ \hline
Step 2: $ \tilde{Q}_m = \sum\nolimits_{k = 1}^K \tilde{p}_{mk}^{\mathrm{MMI-LS}}, \,\, m = 1, \cdots ,M.$ \\ \hline
Sept 3: let $\tilde{Q}_{\max } = \max_{m = 1}^M \tilde{Q}_m$, then $p_{mk}^{\mathrm{MMI-LS}} = \frac{\tilde {p}_{mk}^{\mathrm{MMI-LS}}}{M\tilde{Q}_{\max }}$. \\ \hline
\end{tabular}
\end{table}

A simple example is provided to illustrate the linear scaling method in Fig. $1$, where $K=2$, $M=3$, $\alpha _1 = \alpha _2 = 1/2$, and the shadow region denotes possible points which satisfy (\ref{f3}). The linear scaling solution $\left( x_1 ,x_2 \right)^{\mathrm{LS}} $ is the intersection point of line $x_1  = x_2$ and the right border of the shadow region. Fig. $1$ shows that linear scaling is not optimal because it does not necessarily minimize the distance in (\ref{f21}). However, the border of the shadow region in the first quadrant is the Pareto border, which offers the candidates of the optimal solution. Otherwise, the distance in (\ref{f21}) can be further minimized by increasing the power of each UE simultaneously. 
In Fig. $1$, $( x_1 ,x_2 )^{\mathrm{Opt}}$ denotes the optimal point, which is different from the solution of linear scaling. In summary, the linear scaling method provides a Pareto solution but not necessarily a globally optimal solution.

\ifCLASSOPTIONonecolumn
\begin{figure}[!t]
\centering \includegraphics[width = 1.0\linewidth]{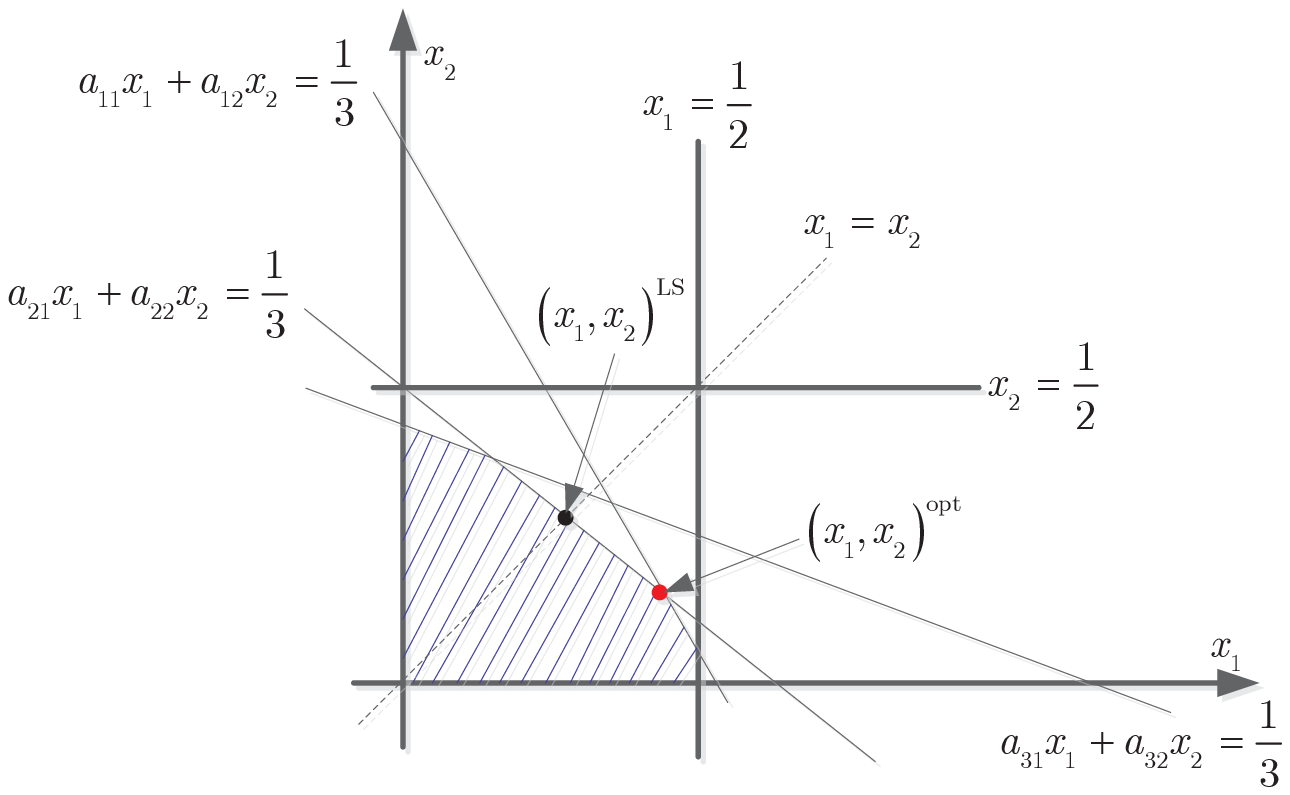}
\caption{ A simple example: $K=2$, $M=3$. }
\end{figure}
\else
\begin{figure}[!t]
\centering \includegraphics[width = 1.0\linewidth]{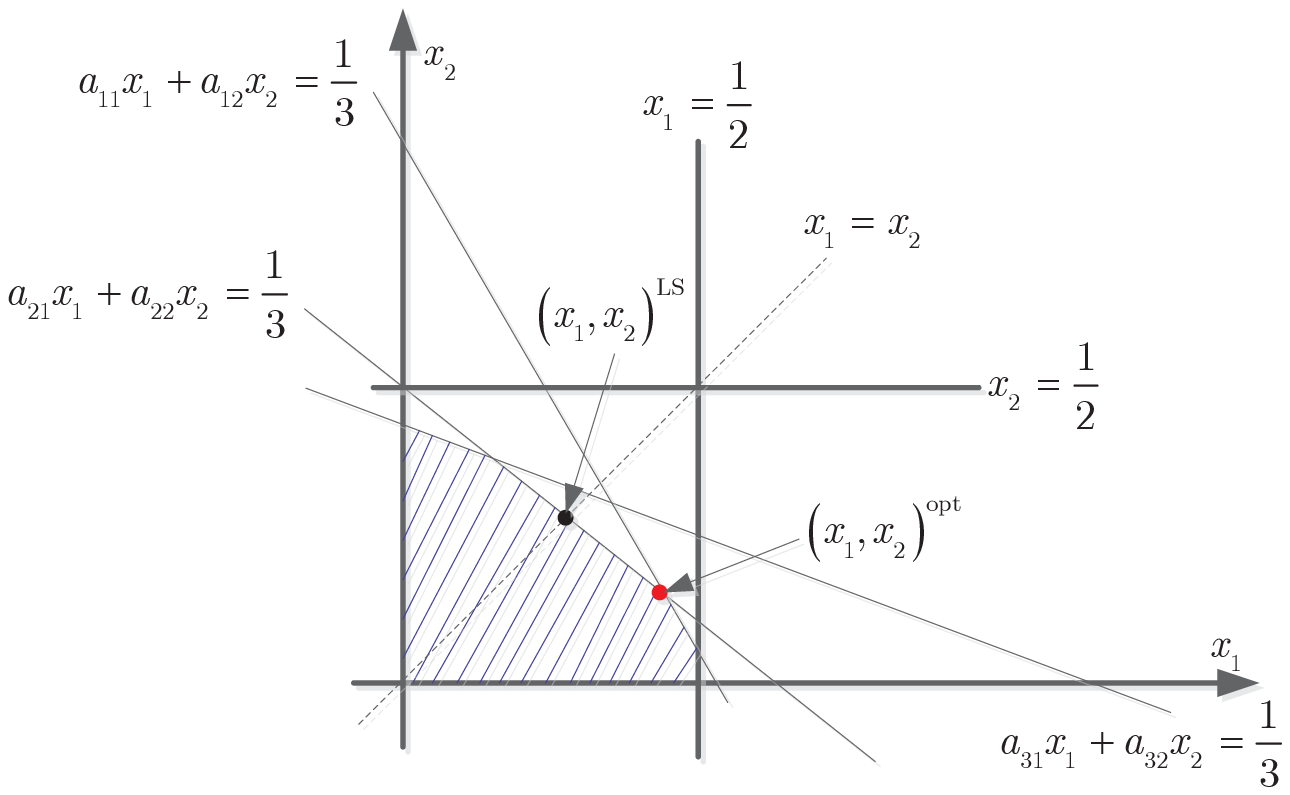}
\caption{ A simple example: $K=2$, $M=3$. }
\end{figure}
\fi

\subsection{Post SINR of the Linear Scaling Method}
Now we provide an approximation of the SINR reduction compared to ZF under the SPC due to the linear scaling of the transmitting power in (\ref{f62}). 
For large-scale MIMO systems where $M$ is very large, based on \cite{Tulino_RMT}, 
$\| \mathbf{H}^{\dag} \|_{\mathrm{F}}^2$ can be approximated as 
\begin{equation}
\label{f23}
\left\| \mathbf{H}^{\dag} \right\|_{\mathrm{F}}^2 = \mathrm{Tr} \left[ \left(\mathbf{H}^{\mathrm{H}} \mathbf{H} \right)^{-1} \right] \approx 
\frac{K}{{M - K}}.
\end{equation}
Since $Q_{\max }$ denotes the maximum squared $2$-norm of the $M$ row vectors of $\mathbf{H}^{\dag}$, it is very hard to derive its expectation directly. Hence, a heuristic approximation is offered here
, 
which is based on two statistical properties of $Q_{\max }$. Firstly, $Q_{\max } \ge K / [( M - K )M ]$ statistically, where $K / [( M - K )M ]$ is the mean squared $2$-norm of rows of $\mathbf{H}^{\dag}$, which can be derived by (\ref{f23}). 
Secondly, $Q_{\max } \to K / [( M - K )M ]$ statistically when $K$ increases from $1$ to $M/2$, which is based on 
the law of large numbers as the elements of $\mathbf{H}^{\dag }$ are approximated as i.i.d. zero-mean complex Gaussian random variables
. In other words, $Q_{\max}$ becomes increasingly larger than $K / [( M - K )M ]$ statistically as $K$ decreases from $M/2$ to $1$. The reason is that as the dimension of the row vectors decrease, the variance of the squared $2$-norm of row vectors of $\mathbf{H}^{\dag}$ increases. 
Note that for large-scale MIMO systems, $M$ is much larger than $K$, so the case of $M/2 < K \leq M$ is ignored. In summary, statistically, ${Q_{\mathrm{max} }}$ should be larger than  $K / [( M - K )M ]$ and approaches to it gradually as $K$ increases from $1$ to $M/2$. Based on the above analysis, we approximate $Q_{\mathrm{max} }$ by multiplying a modifying factor $f_{\mathrm{LS}} ( K,M)$ to $K / [( M - K )M ]$ as
\begin{equation}
\label{f24}
Q_{\mathrm{max}} \approx \frac{Kf_{\mathrm{LS}}\left( K,M \right)}{\left( M - K \right)M},\,\, 1 \leq K \leq \frac{M}{2}
\end{equation}  
where $f_{\mathrm{LS}} ( K,M )$ is empirically selected to be larger than $1$ and approach to $1$ as $K$ increases from $1$ to $M/2$, as $f_{\mathrm{LS}} ( K,M ) = ( M/2K )^p + \beta_{\mathrm{LS}} ( M/2K )^q$ with positive real numbers of $p$, $q$, and $\beta_{LS}$. 
Note that $\beta_{\mathrm{LS}}( \frac{M}{2K} )^q$ ensures $f_{\mathrm{LS}}( K,M )>1$. The values of $p$, $q$, $\beta_{\mathrm{LS}}$ can be chosen by the curve fitting method \cite{Guest_CF}, which are empirically selected as $1/5$, $1/4$, and $1/8$ respectively in this paper. Fig. $2$ illustrates the differences between the real values (Real in the figure) and the approximations (\ref{f24}) (Approx in the figure) with different values of $M$, where $K$ varies from $4$ to $M/2$. It shows that the approximation (\ref{f24}) provides a reasonable estimation of the real value for various $K$ and $M$ values in large-scale MIMO systems. 

\ifCLASSOPTIONonecolumn
\begin{figure}[!t]
\centering \includegraphics[width = 1.0\linewidth]{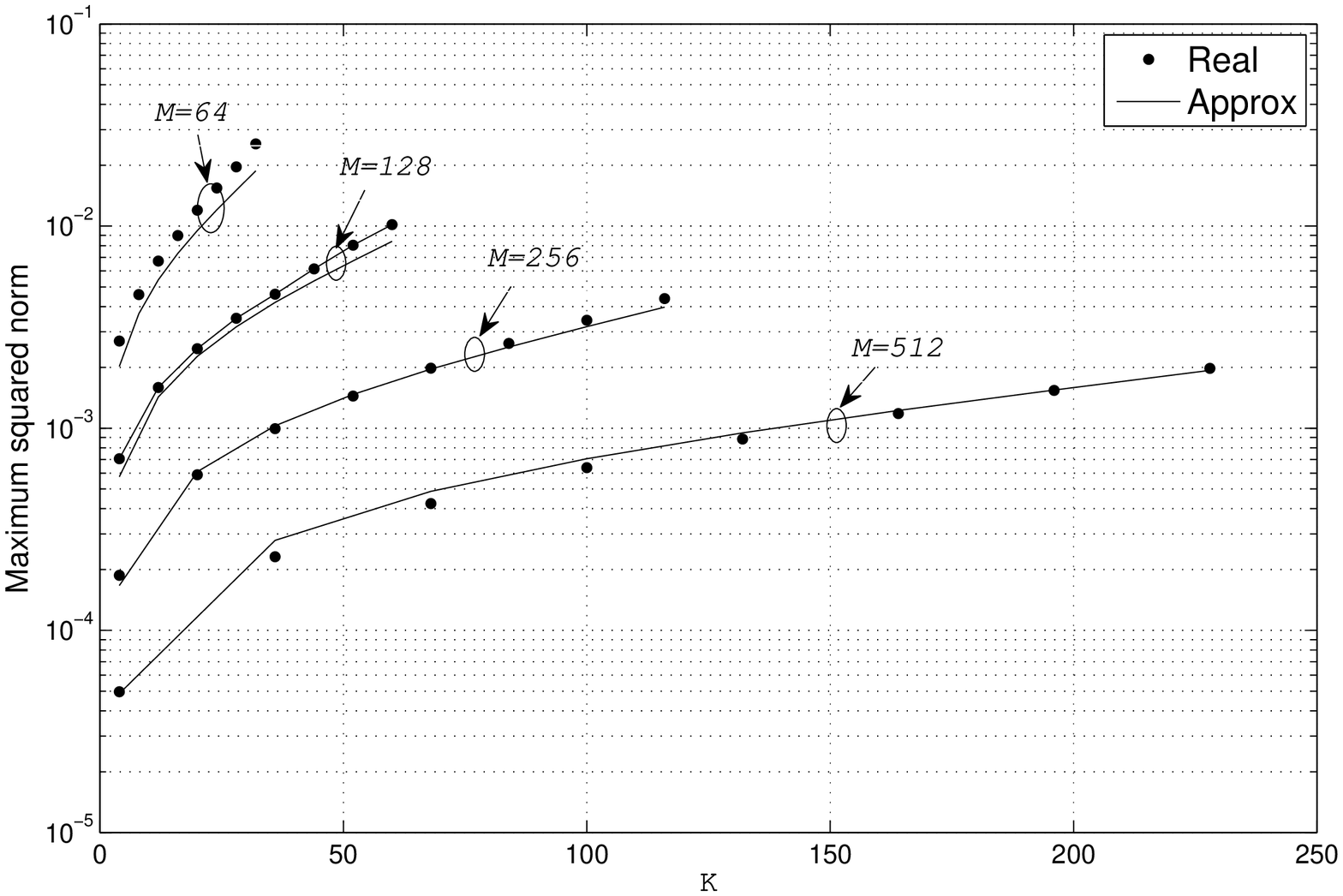}
\caption{Comparison between the real value of $Q_{max}$ and its approximation (\ref{f24}).}
\end{figure}
\else
\begin{figure}[!t]
\centering \includegraphics[width = 1.0\linewidth]{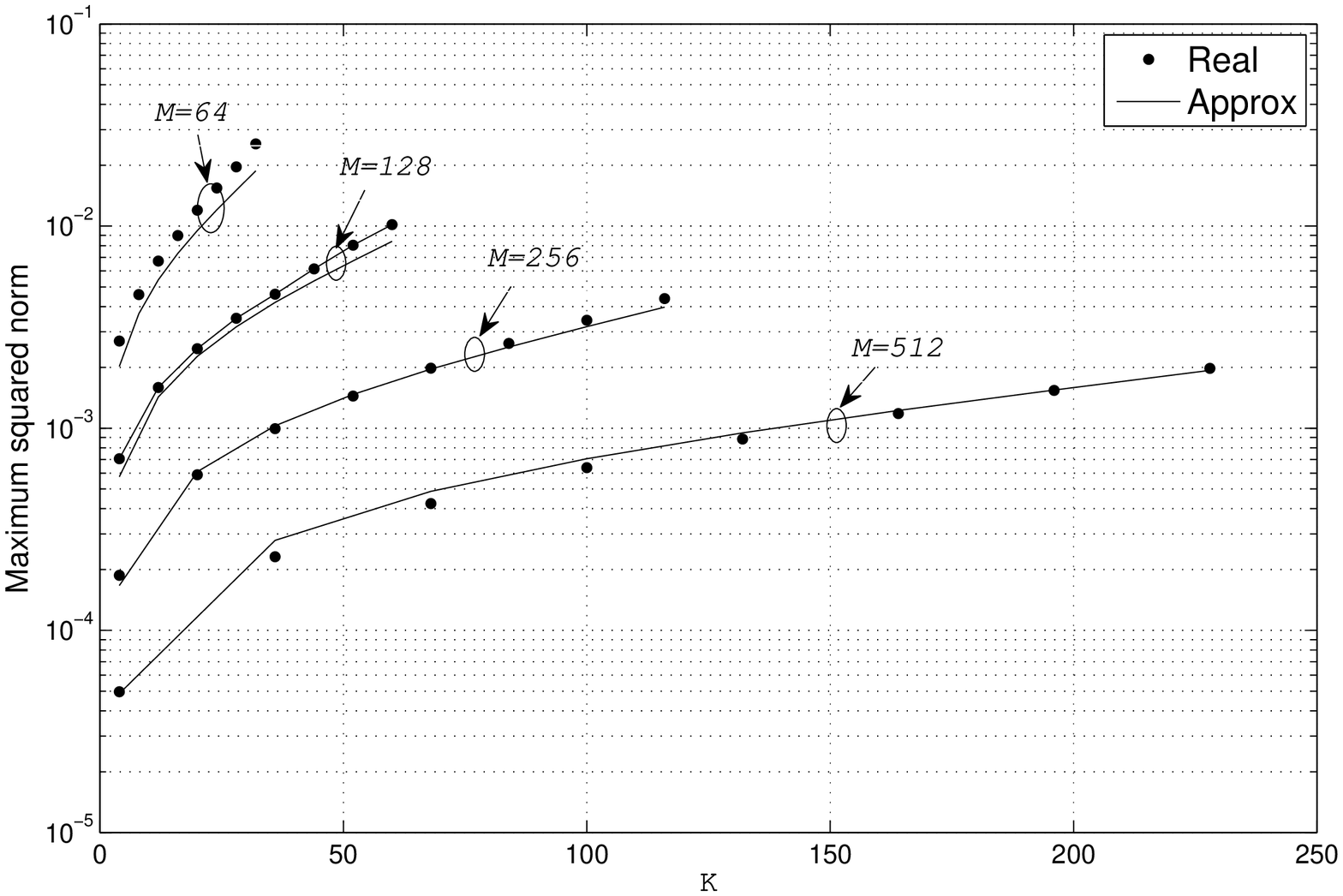}
\caption{Comparison between the real value of $Q_{max}$ and its approximation (\ref{f24}).}
\end{figure}
\fi

Based on (\ref{f62})-(\ref{f24}), the total power allocated to the $k^{th}$ UE from summing all antennas  can be approximated as $x_k^{\mathrm{MMI}} \approx \alpha_k^{\mathrm{ZF-SPC}}/f_{\mathrm{LS}}( K,M )$, hence the ratio of SINR between the virtually optimal and the linear scaling solutions is  
\ifCLASSOPTIONonecolumn
\begin{align}
\label{f25}
G_k^{\mathrm{Opt-LS}} & = \frac{\mathrm {SINR}^{\mathrm{Opt}}}{\mathrm {SINR}^{\mathrm{LS}}} \nonumber \\ 
& \approx \frac{\left| \mathbf{h}^{\mathrm{T}}_k \mathbf{w}^{\mathrm{ZF-SPC}}_k \right|^2}{\sum_{l=1,l\neq k}^K \left| \mathbf{h}^{\mathrm{T}}_k \mathbf{w}^\mathrm{ZF-SPC}_l \right|^2 + \sigma_{\mathrm{n},k}^2} \times \frac{\sum_{l=1, l\neq k}^K \frac{1}{f_{\mathrm{LS}} \left( K,M \right)} \left| \mathbf{h}^{\mathrm{T}}_k \mathbf{w}^\mathrm{ZF-SPC}_l \right|^2 + \sigma _{\mathrm{n},k}^2}{\frac{1}{f_{\mathrm{LS}} \left( K,M \right)} \left| \mathbf{h}^{\mathrm{T}}_k \mathbf{w}^\mathrm{ZF-SPC}_k \right|^2}  \nonumber \\
& = \frac{\sum_{l=1, l\neq k}^K \left| \mathbf{h}^{\mathrm{T}}_k \mathbf{w}^\mathrm{ZF-SPC}_l \right|^2 + f_{\mathrm{LS}} \left( K,M \right)\sigma _{\mathrm{n},k}^2}{\sum_{l=1,l\neq k}^K \left| \mathbf{h}^{\mathrm{T}}_k \mathbf{w}^\mathrm{ZF-SPC}_l \right|^2 + \sigma_{\mathrm{n},k}^2}.
\end{align}
\else
\begin{align}
\label{f25}
G_k^{\mathrm{Opt-LS}} & = \frac{\mathrm {SINR}^{\mathrm{Opt}}}{\mathrm {SINR}^{\mathrm{LS}}} \nonumber \\ 
& \approx \frac{\left| \mathbf{h}^{\mathrm{T}}_k \mathbf{w}^{\mathrm{ZF-SPC}}_k \right|^2}{\sum_{l=1,l\neq k}^K \left| \mathbf{h}^{\mathrm{T}}_k \mathbf{w}^\mathrm{ZF-SPC}_l \right|^2 + \sigma_{\mathrm{n},k}^2} \times \frac{\sum_{l=1, l\neq k}^K \frac{1}{f_{\mathrm{LS}} \left( K,M \right)} \left| \mathbf{h}^{\mathrm{T}}_k \mathbf{w}^\mathrm{ZF-SPC}_l \right|^2 + \sigma _{\mathrm{n},k}^2}{\frac{1}{f_{\mathrm{LS}} \left( K,M \right)} \left| \mathbf{h}^{\mathrm{T}}_k \mathbf{w}^\mathrm{ZF-SPC}_k \right|^2}  \nonumber \\
& = \frac{\sum_{l=1, l\neq k}^K \left| \mathbf{h}^{\mathrm{T}}_k \mathbf{w}^\mathrm{ZF-SPC}_l \right|^2 + f_{\mathrm{LS}} \left( K,M \right)\sigma _{\mathrm{n},k}^2}{\sum_{l=1,l\neq k}^K \left| \mathbf{h}^{\mathrm{T}}_k \mathbf{w}^\mathrm{ZF-SPC}_l \right|^2 + \sigma_{\mathrm{n},k}^2}.
\end{align}
\fi
When the ideal CSI is available at the BS, the multi-user interference $\sum_{l = 1, l \ne k} ^K | \mathbf{h}^{\mathrm{T}}_k \mathbf{w}^{\mathrm{ZF-SPC}}_l |^2$ in (\ref{f25}) is zero, and the SINR ratio is $G_k^{\mathrm{Opt-LS}} \approx f_{\mathrm{LS}}( K,M )$, e.g., about $2\mathrm{dB}$ loss of linear scaling when $K=16$ and $M=128$. Since the SINR of the virtually optimal precoding can be estimated as in \cite{Zhu_Grassmann_arXiv}, the SINR of linear scaling can be approximated 
by
\begin{equation}
\label{f26}
\mathrm {SINR}^{\mathrm{LS}} \approx \frac{{\mathrm {SINR}^{\mathrm{Opt}}}}{{f_{\mathrm{LS}}\left( {K,M} \right)}}.
\end{equation}
When the CSI is non-ideal, the multi-user interference is no longer zero, and (\ref{f25}) can be rewritten as
\begin{equation}
\label{f27}
G_k^{\mathrm{Opt-LS}} \approx f_{\mathrm{LS}}\left( K,M \right) - \frac{f_{\mathrm{LS}}\left( K,M \right) - 1}{1 + \gamma _{\mathrm{N2I},k}}
\end{equation}
where $\gamma _{\mathrm{N2I},k} = \sigma _{\mathrm{n},k}^2 / \sum_{l=1, l \ne k} ^K | \mathbf{h}^{\mathrm{T}}_k \mathbf{w}^{\mathrm{ZF-SPC}}_l |^2$. 
According to \cite{Zhu_Grassmann_arXiv}, $\gamma _{\mathrm{N2I}}$ can be estimated as
\begin{equation}
\label{f28}
\gamma _{\mathrm{N2I},k} \approx \frac{\sigma _{\mathrm{n},k}^2K\left( M - K + 1 \right)}{M\left( 1 - \beta^{2} \right)\left( K - 2 \right)}
\end{equation}
where $\beta$ is the normalized correlation coefficient between the ideal and measured channel vectors, i.e.
\begin{equation}
\label{f70}
\frac{\mathbf{h}_k}{\left\| \mathbf{h}_k \right\|} = \beta \frac{\mathbf{h}_k^{\mathrm{Ideal}}}{\left\| \mathbf{h}_k^{{\mathrm{Ideal}}} \right\|} + \sqrt {1 - \beta ^2} \frac{\mathbf{h}_k^{\mathrm{Ideal}, \bot }}{\left\| \mathbf{h}_k^{\mathrm{Ideal}, \bot } \right\|},
\end{equation}
where $\mathbf{h}_k^{\mathrm{Ideal}}$ denotes the ideal CSI and $\mathbf{h}_k^{\mathrm{Ideal}, \bot }$ denotes the vector in the null space of $\mathbf{h}_k^{\mathrm{Ideal}}$,
 and 
$0\leq \beta\leq1$. Then, $G_k^{\mathrm{Opt-LS}}$ could be estimated by a function of $K$, $M$, $\beta$, and $\sigma_{\mathrm{n},k}$ as
\ifCLASSOPTIONonecolumn
\begin{equation}
\label{f29}
G_k^{\mathrm{Opt-LS}} \approx f_{\mathrm{LS}}\left( K,M \right) - \frac{M\left[ f_{\mathrm{LS}}\left( K,M \right) - 1 \right]\left( 1 - \beta^{2} \right)\left( K - 2 \right)}{M\left( 1 - \beta^{2} \right)\left( K - 2 \right) + K\left( M - K + 1 \right)\sigma _{\mathrm{n},k}^2}.
\end{equation}
\else
\begin{align}
\label{f29}
G_k^{\mathrm{Opt-LS}} \approx f_{\mathrm{LS}}\left( K,M \right) - \frac{M\left[ f_{\mathrm{LS}}\left( K,M \right) - 1 \right]\left( 1 - \beta^{2} \right)\left( K - 2 \right)}{M\left( 1 - \beta^{2} \right)\left( K - 2 \right) + K\left( M - K + 1 \right)\sigma _{\mathrm{n},k}^2}.
\end{align}
\fi
Given $K$ and $M$, when $\beta = 1$, (\ref{f29}) achieves its maximum value $f_{\mathrm{LS}}(K,M)$ regardless of the SNR, which means that linear scaling suffers the largest sum rate loss with ideal CSI. When $\beta < 1$, as SNR increases, i.e., $\sigma_{\mathrm{n},k}^2$ approaches to $0$, (\ref{f29}) approaches to $1$, which means that the sum rate loss of linear scaling is almost $0$ in the relatively high SNR region with non-ideal CSI. As the SNR decreases, i.e., $\sigma_{\mathrm{n},k}^2$ goes to infinity, (\ref{f29}) approaches to the maximum value $f_{\mathrm{LS}}(K,M)$, which means that the gap between the virtually optimal and the linear scaling methods increases as SNR decreases with non-ideal CSI. Given the SNR value, as $\beta$ increases from $0$ to $1$, (\ref{f29}) keeps increasing from $f_{\mathrm{LS}}(K,M)-[Mf_{\mathrm{LS}}(K,M)-1)(K-2)]/[M(K-2)+(KM-K+1)\sigma_{\mathrm{NI},k}^2]$ to $f_{\mathrm{LS}}(K,M)$. In summary, linear scaling works best in the high SNR region with non-ideal CSI, and it is closer to the virtually optimal method as $\beta$ decreases. 
Fig. $3$ provides comparisons between (\ref{f29}) and the real SINR gain with various values of $\beta$ and SNR where $M=256$ and $K=24$, which verifies the analysis above. Moreover, the difference between the real and approximated values in (\ref{f29}) decreases as SNR increases with the maximum error less than $0.2\mathrm{dB}$. Hence, (\ref{f29}) is a reasonable approximation of the SINR gain. Given the values of $M$, $K$, and $\sigma _{\mathrm{n},k}$, with the estimated $\mathrm {SINR}^{\mathrm{Opt}}$ in \cite{Zhu_Grassmann_arXiv}, the sum rate of linear scaling with various values of $\beta$ can be estimated using (\ref{f25}) and (\ref{f29}). 

\ifCLASSOPTIONonecolumn
\begin{figure}[!t]
\centering \includegraphics[width = 1.0\linewidth]{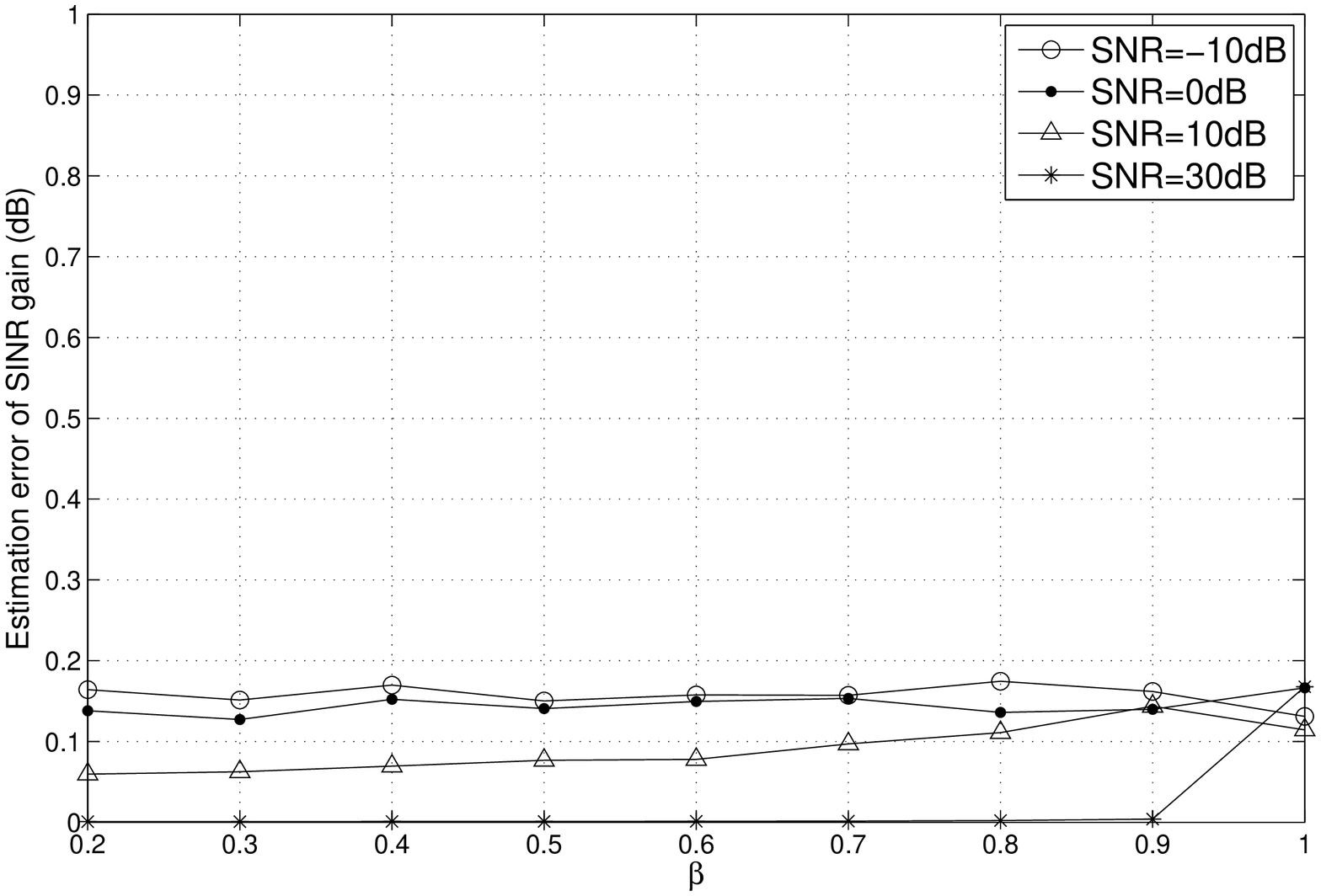}
\caption{  The approximation error of (\ref{f29}), with $M=256$ and $K=24$.}
\end{figure}
\else
\begin{figure}[!t]
\centering \includegraphics[width = 1.0\linewidth]{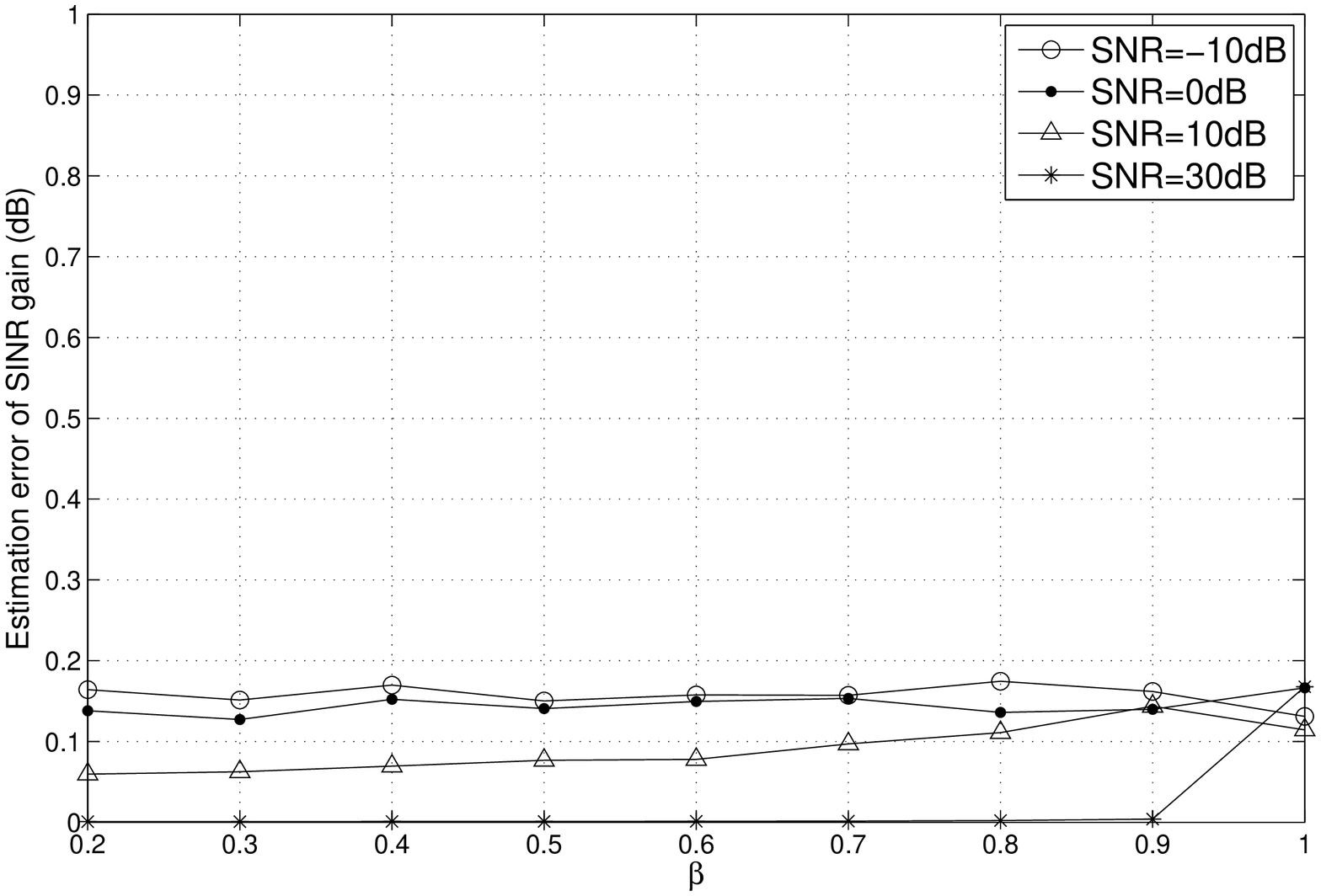}
\caption{The approximation error of (\ref{f29}), with $M=256$ and $K=24$.}
\end{figure}
\fi

\subsection{Newton Iterative Method}
Similar to Subsection III-C, Problem (\ref{f21}) can be solved by iterative search with the interior-point method \cite{Boyd_CO}, where the object function is written as
\begin{equation}
\label{f63}
\begin{split}
&\min f_{\mathrm{MMI}}\left(\mathbf{x}^{\mathrm{MMI}}\right)=  t\left\| \sqrt {\mathbf{p}^{\mathrm{MMI}}}  - \sqrt{\mathbf{r}} \right\|_2^2 - \sum\limits_{i = 1}^{M + K} \log f_i \left( \mathbf{x}^{\mathrm{MMI}} \right), \\
&\mathrm{s.t.\,\,} \mathbf{p}^{\mathrm{MMI}} = \mathrm{vec} \left[ \mathbf{A}^{\mathrm{MMI}} \mathrm{diag} \left( \mathbf{x}^{\mathbf{MMI}} \right) \right],
\end{split}
\end{equation}
where $f_i ( \mathbf{x}^{\mathrm{MMI}} ) = \tilde{\mathbf{a}}^{\mathrm{T}}_i \mathbf{x}^{\mathrm{MMI}} - b_i$, $i=1,\cdots, M+K$, and $\tilde{\mathbf{a}}_i^T $ is the $i$th row vector of the matrix $\tilde{\mathbf{A}}^{\mathrm{MMI}} = [ \mathbf{A}^{\mathrm{MMI,T}}, \,\,\mathbf{I}_K ]^\mathrm{T}$. With the Newton iterative method \cite{Boyd_CO}, the gradient vector and Hessian matrix of (\ref{f63}) is calculated as 
\begin{equation}
\label{f64}
\nabla f_{\mathrm{MMI}} = t\sum\limits_{i = 1}^M \left( \tilde{\mathbf{a}}_i^{\mathrm{T}} - \frac{\sqrt{\tilde{\mathbf{a}}_i^{\mathrm{T}}}} { \sqrt {\mathbf{x}^{\mathrm{MMI}}} }\circ \sqrt{\mathbf{p}_i^{\mathrm{ZF-SPC,T}}} \right) + \sum\limits_{i = 1}^{M + K} \frac{ - \tilde{\mathbf{a}}_i^{\mathrm{T}}}{\tilde{\mathbf{a}}_i^{\mathrm{T}}\mathbf{x}^{\mathrm{MMI}} - b_i}  
\end{equation}
and
\begin{align}
\label{f65}
\nabla ^2 f_{\mathrm{MMI}} & = \left[ \begin{array}{*{20}{c}}
 - \sum\limits_{m = 1}^M \left( a_{m1}^{\mathrm{MMI}} p_{m1}^{\mathrm{ZF-SPC}} \right)^{\frac{1}{2}} \left( x_1^{\mathrm{MMI}} \right)^{-\frac{3}{2}} & & \\
 & \ddots & \\
 & & - \sum\limits_{m = 1}^M \left( a_{mK}^{\mathrm{MMI}} p_{mK}^{\mathrm{ZF-SPC}} \right)^{\frac{1}{2}} \left( x_K^{\mathrm{MMI}} \right)^{-\frac{3}{2}} \end{array} \right] \nonumber \\
 & + \sum\limits_{i = 1}^{M + K} \frac{ - \tilde{\mathbf{a}}_i^{\mathrm{T}} \tilde{\mathbf{a}}_i}{\left( \tilde{\mathbf{a}}_i^{\mathrm{T}} \mathbf{x}^{\mathrm{MMI}} - b_i \right)^2} 
\end{align}
respectively. Then, the Newton iteration method can be carried out as in \cite{Boyd_CO}. 

\section{Computation Complexity}
In this section, we analyze the computation complexity of the four discussed ZF-based power allocation algorithms for large-scale MIMO systems under the PAPC. 
\subsection{Orthogonal Projection Method}
As for the orthogonal projection algorithm proposed in Section III-A, most of the computation is introduced by (\ref{f13}) and (\ref{f14}).
Let $\tilde{\mathbf{r}}= [ \tilde{\mathbf{r}}_1^{\mathrm{T}}, \,\, \tilde{\mathbf{r}}_2^{\mathrm{T}}, \cdots, \tilde{\mathbf{r}}_K^{\mathrm{T}}, \,\, \tilde{r} ]^{\mathrm{T}}$ and $\tilde{\mathbf{p}}^{\mathrm{Sub}} = [\tilde{\mathbf{p}}_1^{\mathrm{T}}, \,\, \tilde{\mathbf{p}}_2^{\mathrm{T}}, \cdots,  \tilde{\mathbf{p}}_K^{\mathrm{T}}, \,\, \tilde{p}]^{\mathrm{T}}$ respectively, where both $\tilde{\mathbf{ r}}_i$ and $\tilde{\mathbf{p}}_i$, $i=1,\cdots,K$, are $M\times 1$ vectors and $\tilde r$ and $\tilde p$ in (\ref{f13}) are scalars. Since the projection matrix $\mathbf{M}^{\mathrm{Proj}}$ has a specific form as in (\ref{f56}), $\tilde{\mathbf{p}}^{\mathrm{Sub}}$ is computed as
\begin{equation}
\label{f36}
\tilde{\mathbf{p}}_i  = \mathbf{D}_1 \tilde{\mathbf{r}}_i  + \mathbf{D}_2 \sum\limits_{k = 1, k \ne i} ^K \tilde{\mathbf{r}}_k + \left(\frac{1}{M^2K^2} - \frac{1}{MK}\right)\tilde {r}, \,\, i = 1, \cdots ,K
\end{equation}
and 
\begin{equation}
\label{f37}
p = \left(\frac{1}{M^2K^2} - \frac{1}{MK}\right) \sum\limits_{k=1}^{K}\sum\limits_{m=1}^{M} \tilde{r}_{mk}+c{\tilde r}.
\end{equation}
In (\ref{f36}), $\sum_{k = 1, k \ne i} ^K \tilde{\mathbf{r}}_k$ could be rewritten as $\sum_{k=1}^{K}\tilde{\mathbf{r}}_k-\tilde{\mathbf{r}}_i$, which needs $2MK-M$ real-valued additions for all the $K$ possible values of $i$, where $MK-M$ of them are from $\sum_{k=1}^{K}\tilde{\mathbf{r}}_k$ and the other $MK$ of them are from the $K$ vector subtractions. With the special form of matrices $\mathbf{D}_1$ and $\mathbf{D}_2$, the $2K$ products $\mathbf{D}_1 \tilde{\mathbf{r}}_i$ and $\mathbf{D}_2 \sum_{k = 1, k \ne i} ^K \tilde{\mathbf{r}}_k$ require $4MK$ real-valued multiplications and $4MK-2K$ real-valued additions. In addition, the rest computation of (\ref{f36}) includes $2MK$ real-valued additions. Hence, 
the computations of (\ref{f36}) involves about $4MK$ real-valued multiplications and $8MK$ real-valued additions. As for (\ref{f37}), it involves $MK$ real-valued additions and $2$ real-valued multiplications. Furthermore, (\ref{f14}) includes $MK$ real-valued divisions, hence its complexity is $O(MK)$. In summary, the total computation complexity of the orthogonal projection method includes $O(MK)$ real-valued multiplications and additions respectively.

\subsection{Feasible Newton Iterative Method}
As for the feasible newton iterative method discussed in Section III-C, most of the computation is caused by solving the Newton decrement equation, where the inverse matrix with the dimension of $MK+M$ has to be computed firstly. Let 
\begin{equation}
\label{f38}
\left[ \begin{array}{*{20}c}
   \mathbf{P} & \mathbf{Q}  \\
   \mathbf{Q}^{\mathrm{T}} & \mathbf{S}  \\

 \end{array}  \right] = \left[ \begin{array}{*{20}c}
   \nabla ^2 f_\mathbf{x} & \mathbf{A}^{\mathrm{T}}  \\
   \mathbf{A} & \mathbf{0}_M  \\
 \end{array}  \right]^{- 1}, 
\end{equation}
where the dimensions of matrices $\mathbf P$, $\mathbf Q$, and $\mathbf S$ are $MK\times MK$, $MK\times M$, and $M\times M$ respectively and $\mathbf{0}_M$ denotes a $M \times M$ all-zero matrix. Based on \cite{Boyd_CO}, only $\mathbf{P}$ needs to be obtained instead of the whole inverse matrix in (\ref{f38}). According to the inverse of block matrix formula \cite{Horn_Matrix_Analysis}, $\mathbf{P}$ could be written as
\begin{equation}
\label{f39}
\mathbf{P} = \left( \nabla ^2 f_{\mathbf{x}} \right)^{ - 1}  + \left( \nabla ^2 f_{\mathbf{x}}  \right)^{ - 1} \mathbf{A}^{\mathrm{T}} \left[ \mathbf{A}\left( \nabla ^2 f_{\mathbf{x}} \right)^{ - 1} \mathbf{A}^{\mathrm{T}} \right]^{ - 1} \mathbf{A}\left( \nabla ^2 f_{\mathbf{x}} \right)^{ - 1}.
\end{equation}
As $\nabla ^2f_{\mathbf{x}}$ is a diagonal matrix, its inverse only involves $O(MK)$ real-valued multiplications and additions respectively. Moreover, as $\mathbf{A} = [\mathbf{I}_M, \,\, \cdots, \,\, \mathbf{I}_M ]$, the computation of $\mathbf{A}^{\mathrm{T}} [ \mathbf{A}( \nabla ^2 f_{\mathbf{x}} )^{-1} \mathbf{A}^{\mathrm{T}} ] \mathbf{A}$ only needs $O(MK)$ multiplications and additions respectively, where the result has a form of 
\begin{equation}
\label{f69}
\mathbf{M}^{\mathrm{FNI}} = \mathbf{A}^{\mathrm{T}} \left[ \mathbf{A} \left( \nabla ^2 f_{\mathbf{x}} \right)^{ - 1} \mathbf{A}^{\mathrm{T}} \right]\mathbf{A} = \left[ \begin{array}{*{20}{c}}
\mathbf{D}_{\mathrm{p}}& \cdots &\mathbf{D}_{\mathrm{p}}\\
 \vdots & \ddots & \vdots \\
\mathbf{D}_{\mathrm{p}}& \cdots &\mathbf{D}_{\mathrm{p}}
\end{array} \right]
\end{equation}
and $\mathbf{D}_{\mathrm{p}}$ is a $M\times M$ diagonal matrix. Hence, the Newton increment $\Delta x_{\mathrm{nt}}$ can be written as
\begin{equation}
\label{70}
\Delta x_{\mathrm{nt}} = \mathbf{P} \nabla f_{\mathbf{x}} = \mathbf{u}_{\nabla f_{\mathbf{x}}} + \left( \nabla ^2 f_{\mathbf{x}} \right)^{ - 1} \left( \mathbf{M}^{\mathrm{FNI}} \mathbf{u}_{\nabla f_{\mathbf{x}}} \right),
\end{equation}
where $\mathbf{u}_{\nabla f_{\mathbf{x}}} = ( \nabla ^2 f_{\mathbf{x}} )^{ - 1} \nabla f_{\mathbf{x}}$ is a $MK\times 1$ vector. Due to the special structure of $\mathbf{M}^{\mathrm{FNI}}$, $\mathbf{M}^{\mathrm{FNI}}\mathbf{u}_{\nabla f_{\mathbf{x}}}$ can be calculated similarly to (\ref{f36}). Hence, $O(MK)$ real-valued multiplications and additions respectively are needed to update $\Delta {x_{nt}}$.
In addition, updating $\mathbf{x}$ and the residual error involve about $2MK$ real-valued multiplications and additions respectively. In summary, the computation complexity of feasible Newton iterative method involves $O(m_{\mathrm{FNI}}MK)$ real-valued multiplications and additions respectively, where $m_{\mathrm{FNI}}$ represents the number of iterations and it is not more than $10$ according to our experiments.



\subsection{Linear Scaling Method}
As for the linear scaling method discussed in Section IV-A, since it only involves simple real-valued multiplications and additions, 
its computation complexity includes $O(MK)$ real-valued multiplications and additions respectively.

\subsection{Newton Iterative Method}
As for the Newton iterative method discussed in Section IV-C, most of the computations are cause by calculating the matrix $\mathbf{A}^{\mathrm{MMI}}$ in (\ref{f19}), the gradient vector $\nabla f_{\mathrm{MMI}}$ in (\ref{f64}), the Hessian matrix $\nabla^2 f_{\mathrm{MMI}}$ in (\ref{f65}), and the Newton decrements. For the matrix $\mathbf{A}^{\mathrm{MMI}}$, it needs about $3MK$ real-valued multiplications and $2MK$ real-valued additions. For the gradient vectors, $\sum_{i = 1}^{M + K} - \tilde{\mathbf{a}}^{\mathrm{T}}_i / (\tilde{\mathbf{a}}_i^{\mathrm{T}} \mathbf{x}^{\mathrm{MMI}} - b_i )$ results in most of the computations and it needs about $O(MK)$ real-valued multiplications and $O(MK)$ real-valued additions as each of the last $K$ vectors of $\tilde{\mathbf {a}}_i$, $i=M+1,\cdots,M+K$, has only one non-zero element. For the Hessian matrix,  $\sum_{i = 1}^{M + K} - \mathbf{a}^{\mathrm{T}}_i \tilde{\mathbf{a}} / (\tilde{\mathbf{a}}_i^{\mathrm{T}} \mathbf{x}^{\mathrm{MMI}} - b_i )$ causes most of the computations and it needs about ${{M{K^2}} \mathord{\left/ {\vphantom {{M{K^2}} 2}} \right.\kern-\nulldelimiterspace} 2} + {{MK} \mathord{\left/{\vphantom {{MK} 2}} \right.\kern-\nulldelimiterspace} 2}$ real-valued multiplications. For the Newton decrements, as $\nabla^2 f_{\mathrm{MMI}}$ is a $K\times K$ matrix and $\nabla f_{\mathrm{MMI}}$ is a $K\times 1$ vector, the inverse of $\nabla^2 f_{\mathrm{MMI}}$, $(\nabla^2 f_{\mathrm{MMI}})^{ - 1} \nabla f_{\mathrm{MMI}}$, and $(\nabla f_{\mathrm{MMI}})^{\mathrm{T}} (\nabla^2 f_{\mathrm{MMI}})^{ - 1} \nabla f_{\mathrm{MMI}}$ need about $O(K^3)$, $K^2$, and $K$ real-valued multiplications and  real-valued additions respectively \cite{Matrix_Computation}. In summary, the total complexity of the Newton iterative Method includes about $O[m_{\mathrm{NI}}(MK^2+K^3)]$ real-valued multiplications and real-valued additions respectively, where $m_{\mathrm{NI}}$ represents the number of iterations and it is not more than $15$ based on our experiments.

\subsection{Summary}
The complexity of the four proposed algorithms are summarized in Table II, where multiplications and additions are real-valued. 
Moreover, for comparison, the computation complexity of the water-filling based solution to Problem (\ref{f66}) is also provided, which is computed based on \cite{Boccardi_ZF_PAPC,Ohwatari_PAPC}. Note that the water-filling based method employed in \cite{Boccardi_ZF_PAPC,Ohwatari_PAPC} is the interior-point method, which has the the complexity similar to our method discussed in Section IV-C, and $m_{\mathrm{WF}}$ denotes its number of iterations. Table II shows that the Newton iterative solution proposed in Section IV-C to Problem (\ref{f21})  and the water-filling based solution employed in \cite{Boccardi_ZF_PAPC,Ohwatari_PAPC} to Problem (\ref{f66}) have the most complexity, while the orthogonal projection solution proposed in Section III-A to Problem (\ref{f11}) and the linear scaling solution proposed in Section IV-A to Problem (\ref{f21}) has the least complexity. The feasible Newton iterative solution proposed in Section III-C to Problem (\ref{f11}) has the medium complexity after the computation simplification discussed in section V-B.   
\begin{table}[!t]
\centering
\renewcommand{\arraystretch}{1.2}
\renewcommand{\captionlabeldelim}{.}
\caption{Comparison of Computation Complexity.}
\begin{tabular}{|c|c|c|c|}
\hline
{Number} &{Multiplications} & {Additions}&{Iterations}\\ \hline
   {Orthogonal projection}&$O(MK)$&$O(MK)$&$1$\\ \hline
 Feasible Newton iteration &$O(m_{\mathrm{FNI}}MK)$&$O(m_{\mathrm{FNI}}MK)$&$m_{\mathrm{FNI}}\leq 10$\\ \hline
   Linear scaling &$O(MK)$&$O(MK)$&$1$\\ \hline
   Newton iteration &$O[m_{\mathrm{NI}}(MK^2+K^3)]$&$O[m_{\mathrm{NI}}(MK^2+K^3)]$&$m_{\mathrm{NI}}\leq 15$\\ \hline
    Water-filling &$O[m_{\mathrm{WF}}(MK^2+K^3)]$&$O[m_{\mathrm{WF}}(MK^2+K^3)]$&$m_{\mathrm{WF}}\leq 15$\\ \hline
\end{tabular}
\end{table}

\section{Conjugate Beamforming}
For the CB precoding, we provide a simple power allocation method which satisfies the PAPC. Let $\mathbf{h}_k=\left|\mathbf{h}_k\right|\circ e^{\boldsymbol{\theta}_k}, k=1,\cdots,K$, where $\boldsymbol{\theta}_k$ is the phase vector of $\mathbf{h}_k$
. Then, the proposed precoding vector of the $k$th UE under the PAPC 
is  
\begin{equation}
\label{f42}
\mathbf{w}^{\mathrm{CB-PAPC}}_k = \sqrt{\frac{\alpha_k^{\mathrm{CB}-SPC}}{M}} e^{-j{\boldsymbol{\theta }_k}}
\end{equation}
where $\alpha_k^{\mathrm{CB-SPC}}$ is the total power allocated to the $k^{th}$ UE from summing all antennas for the CB precoding under the SPC, which is uniformly distributed to each antenna for the PAPC. Note that the CB precoding matrix is provided in \cite{Rusek_massive_MIMO_overview} as $\mathbf{w}^{\mathrm{CB-SPC}}_k = \sqrt{\phi^{\mathrm{CB-SPC}}} \mathbf{h}^{\mathrm{H}} $ where $\phi^{\mathrm{CB-SPC}} = 1 / \mathrm{Tr} [( \mathbf{H} \mathbf{H}^{\mathrm{H}} )]$ with the total maximal power being assumed to be $1$ in this paper. With (\ref{f42}), the power of each antenna is fully used
. 

The post-receiving SINR of the $k$th UE is
\begin{equation}
\label{f43}
\mathrm {SINR}^{\mathrm{CB-PAPC}}_k = \frac{\left| \mathbf{h}_k^{\mathrm{Ideal,T}}\mathbf{w}^{\mathrm{CB-PAPC}}_k \right|^2}{\sum_{l = 1, l \ne k} ^K \left| \mathbf{h}_k^{\mathrm{Ideal,T}} \mathbf{w}^{\mathrm{CB-PAPC}}_l \right|^2  + \sigma _{\mathrm{n},k}^2} ,
\end{equation}
With (\ref{f70}), substituting (\ref{f42}) into (\ref{f43}), it becomes
\begin{equation}
\label{f44}
\mathrm {SINR}^{\mathrm{CB-PAPC}}_k = \frac{\alpha_k^{\mathrm{CB-SPC}}\beta^2 \left( \sum\nolimits_{m = 1}^M \left| h_{km}^{\mathrm{Ideal}} \right| \right)^2}{M\left\| \mathbf{h}^{\mathrm{Ideal}}_k \right\|^2 \sum\nolimits_{l = 1,l \ne k}^K \alpha _l^{\mathrm{CB-SPC}}\left| \mathbf{v}^{\mathrm{T}}_k \mathbf{u}_l \right|^2 + M\sigma _{\mathrm{n},k}^2}
\end{equation}
where $\mathbf{v}_k = \mathbf{h}_k^{\mathrm{Ideal}} / \| \mathbf{h}_k^{\mathrm{Ideal}}\| $ and $\mathbf{u}_l = e^{ - j\boldsymbol{\theta}_l} / \sqrt{M} $. Note that $\mathbf{v}_k$ and $\mathbf{u}_l$ are elements of Grassmann manifold $\mathbb{G} ( 1,M )$ \cite{Zhu_Grassmann_arXiv}. As the elements of $\mathbf{h}_k^{\mathrm{Ideal}}$, $k=1,\cdots,K$, are i.i.d. normalized complex Gaussian random variables, their amplitudes obey Rayleigh distribution. Then, $\sum_{m = 1}^M |h_{km}^{\mathrm{Ideal}} | \to M \mathrm{E} ( | h_{km}^{\mathrm{Ideal}} | ) = M\sqrt{\pi}/2$ when $M$ is large \cite{Proakis_DC,Madhow_ComFund}. For the SPC case, the post-receiving SINR could be written as
\begin{equation}
\label{f45}
\begin{split} 
\mathrm{SINR}_k^{\mathrm{CB-SPC}} & = \frac{\alpha _k^{\mathrm{CB-SPC}} \beta^2 \left\| \mathbf{h}_k^{\mathrm{Ideal}} \right\|^2} {\sum\nolimits_{l = 1,l \ne k}^K \alpha _l^{\mathrm{CB-SPC}} \left\|\mathbf{h}_k^{\mathrm{Ideal,T}} \mathbf{u}_l \right\|^2 + \sigma _{\mathrm{n},k}^2} \\
 & = \frac{\alpha _k^{\mathrm{CB-SPC}} \left\| \mathbf{h}_k^{\mathrm{Ideal}} \right\|^4}{\left\| \mathbf{h}^{\mathrm{Ideal}}_k \right\|^4 \sum\nolimits_{l = 1,l \ne k}^K \alpha _l^{\mathrm{CB-SPC}} \left| \mathbf{v}^{\mathrm{T}}_k \mathbf{u}_l \right|^2 + \left\| \mathbf{h}_k^{\mathrm{Ideal}} \right\|^2 \sigma _{\mathrm{n},k}^2}
\end{split}
\end{equation}
As $M$ is large, $\| \mathbf{h}^{\mathrm{Ideal}}_k \|^2 \approx M$ according to the law of large numbers. Since $\mathbf{v}_k,\mathbf{u}_l \in \mathbb {G} ( 1,M )$, the mutual interference terms $\sum_{l = 1,l  \ne k} ^K |\mathbf{v}^{\mathrm{T}}_k \mathbf{u}_l |^2 $ in (\ref{f44}) and (\ref{f45}) approach to the same value $(K - 1)/M$ \cite{Zhu_Grassmann_arXiv}. Hence, the ratio between $\mathrm {SINR}_k^{\mathrm{CB-PAPC}}$ and $\mathrm {SINR}_k^{\mathrm{CB-SPC}}$ is 
\begin{equation}
\label{f46}
G^{\mathrm{SPC-PAPC}} = \frac{\mathrm {SINR}_k^{\mathrm{CB-SPC}}}{\mathrm {SINR}_k^{\mathrm{CB-PAPC}}} \approx \frac {\left\| \mathbf{h}_k^{\mathrm{Ideal}}\right\|^4}{\left( \sum\nolimits_{m = 1}^M \left| h_{km}^{\mathrm{Ideal}} \right| \right)^2}  \approx \frac{4}{\pi}.
\end{equation}
The approximation (\ref{f46}) implies that the rate loss due to the PAPC of CB 
is $ \log _2 ( 4/\pi ) = 0.3485 \mathrm{bits/s/Hz}$ for each UE compared to the SPC and independent of CSI error, when $\mathrm{SINR}_k^{\mathrm{CB-SPC}}$ and $\mathrm{SINR}_k^{\mathrm{CB-PAPC}}$ are much larger than $1$, which is the preferable case for large-scale MIMO systems. Fig. $4$ compares the real SINR gain and the estimated gain computed from (\ref{f46}) with various values of SNR and $\beta$, which verifies the rationality of (\ref{f46}), as the maximal estimation error dose not exceed $0.2\mathrm{dB}$. Hence, (\ref{f46}) is a reasonable approximation of the SINR gain. 
With the estimated $\mathrm {SINR}^{\mathrm{CB-SPC}}$ in \cite{Zhu_Grassmann_arXiv}, the sum rate of CB precoding under the PAPC could be estimated using (\ref{f46}). 
\ifCLASSOPTIONonecolumn
\begin{figure}[!t]
\centering \includegraphics[width = 1.0\linewidth]{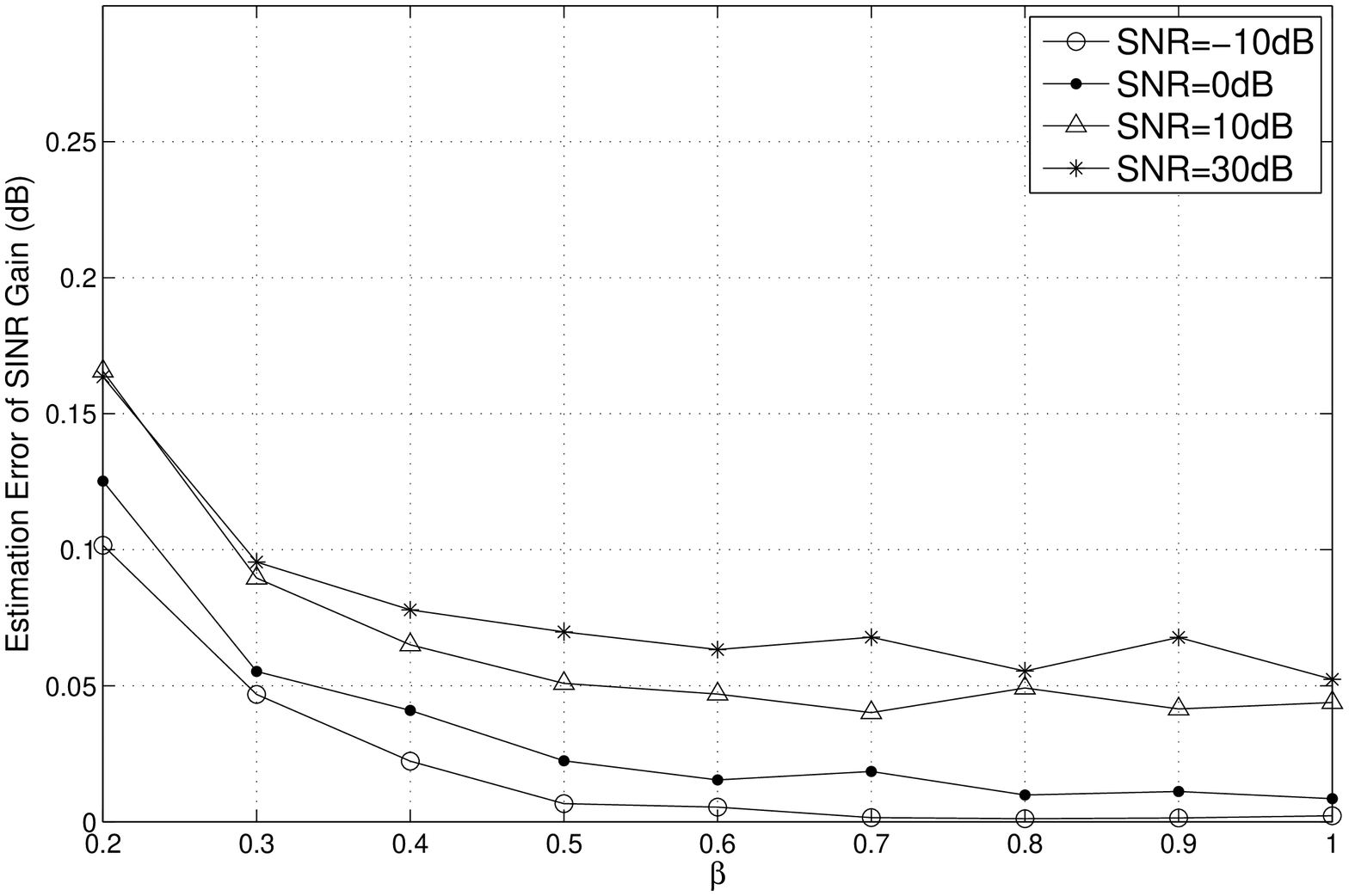}
\caption{The approximation error of (\ref{f46}) with $M=256$ and $K=24$.}
\end{figure}
\else
\begin{figure}[!t]
\centering \includegraphics[width = 1.0\linewidth]{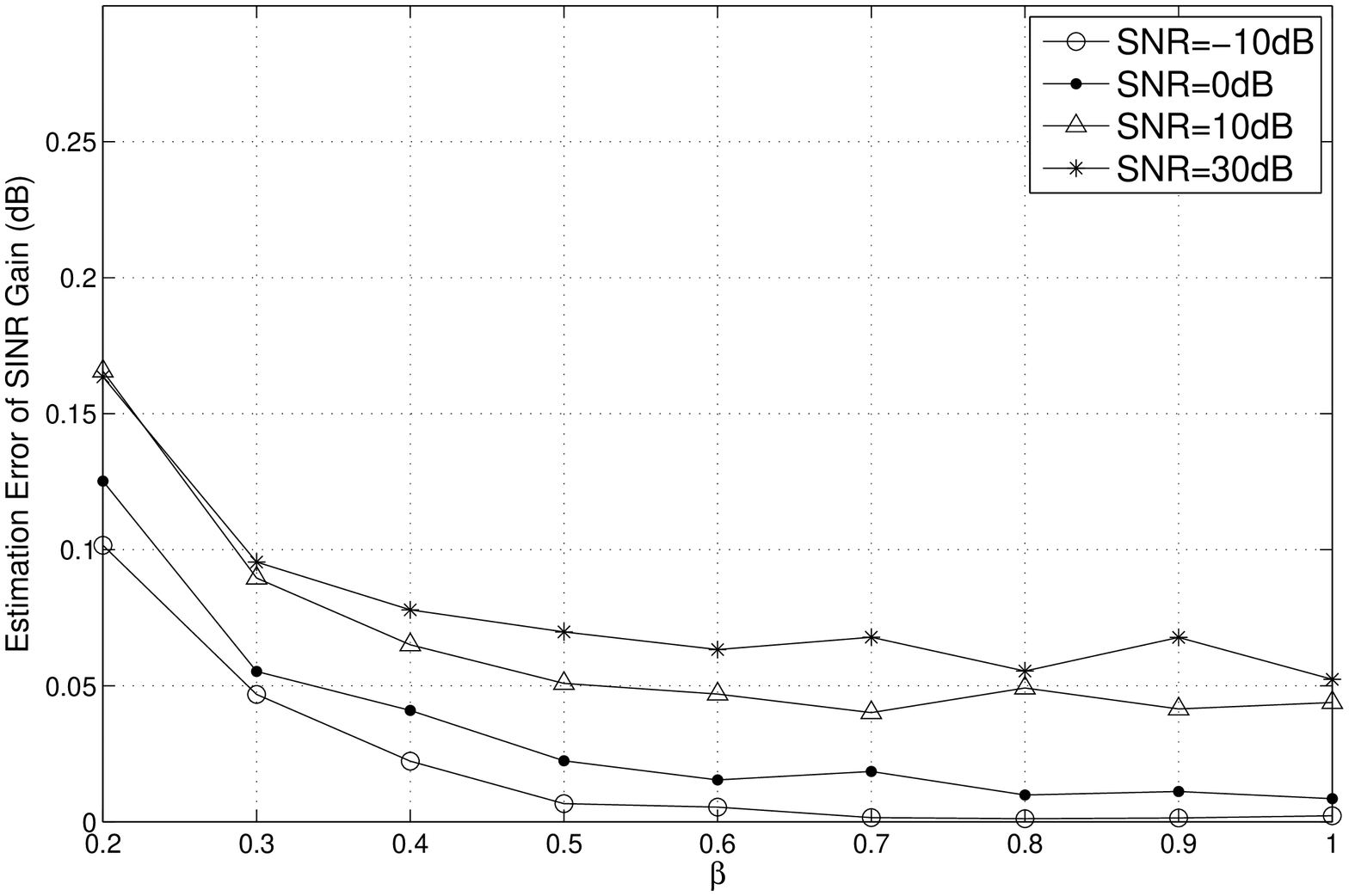}
\caption{The approximation error of (\ref{f46}) with $M=256$ and $K=24$.}
\end{figure}
\fi

\section{Simulation Results}
In this section, simulation results are presented to compare the four ZF-based and one CB-based power allocation algorithms proposed in this paper, where MPU-Proj-ZF, MPU-Opt-ZF, MMI-LS-ZF, and MMI-Opt-ZF denote the orthogonal project, feasible Newton iterative, linear scaling, and Newton iterative methods for the ZF-based power allocation respectively, while PAPC-CB denotes the CB-based power allocation. 
In addition, the approximation of linear scaling, denoted by Est-LS-ZF, introduced in Section IV-B is also provided. For reference, the results of the virtually optimal ZF under the SPC and the simple CB under the SPC, denoted by SPC-ZF and SPC-CB respectively, are provided. Note that for SPC-ZF and SPC-CB, the corresponding precoding matrices employed in \cite{Rusek_massive_MIMO_overview} are used to 
compute the power allocation matrices $\mathbf{P}^{\mathrm{ZF-SPC}}$ and $\mathbf{P}^{\mathrm{CB-SPC}}$ respectively, and the results are used for the power allocation methods proposed in this paper. In addition, the water-filing based solution to Problem (\ref{f66}) employed in \cite{Boccardi_ZF_PAPC,Ohwatari_PAPC}, denoted WF-ZF, is also shown in the figures for reference.

\subsection{Ideal CSI}
In order to investigate the performance with ideal CSI, the simulation results of $K=16, M=128$ and $K=24, M=256$ are provided in Fig. $5$ and Fig. $6$ respectively, where the total maximal transmitting power is assumed to be $1$, with an equal transmitting power constraint $1/M$ of each antenna except for SPC-ZF and SPC-CB.


\ifCLASSOPTIONonecolumn
\begin{figure}[!t]
\centering \includegraphics[width = 1.0\linewidth]{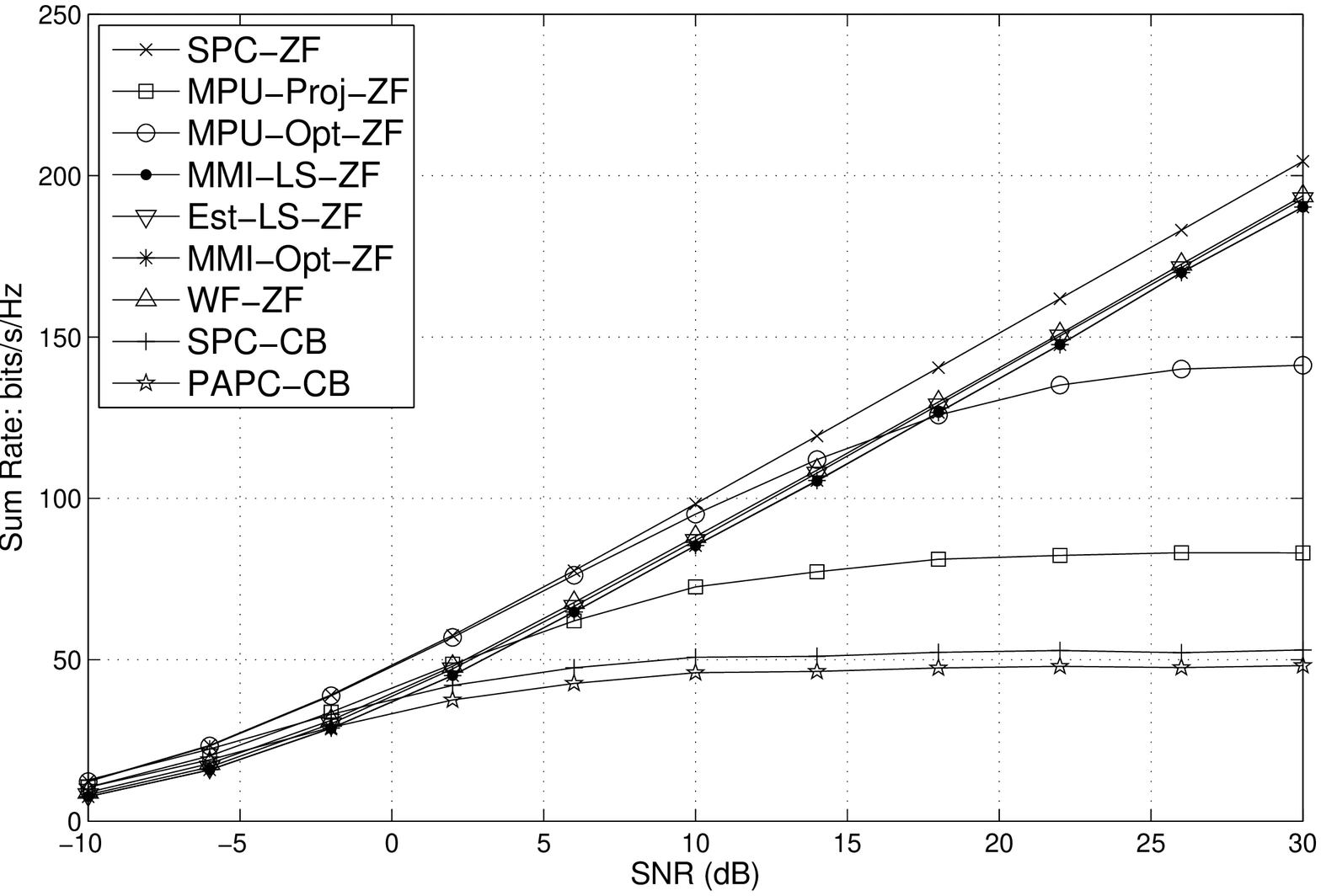}
\caption{Ideal CSI, sum rate vs. SNR, $K=16,M=128$.}
\end{figure}
\else
\begin{figure}[!t]
\centering \includegraphics[width = 1.0\linewidth]{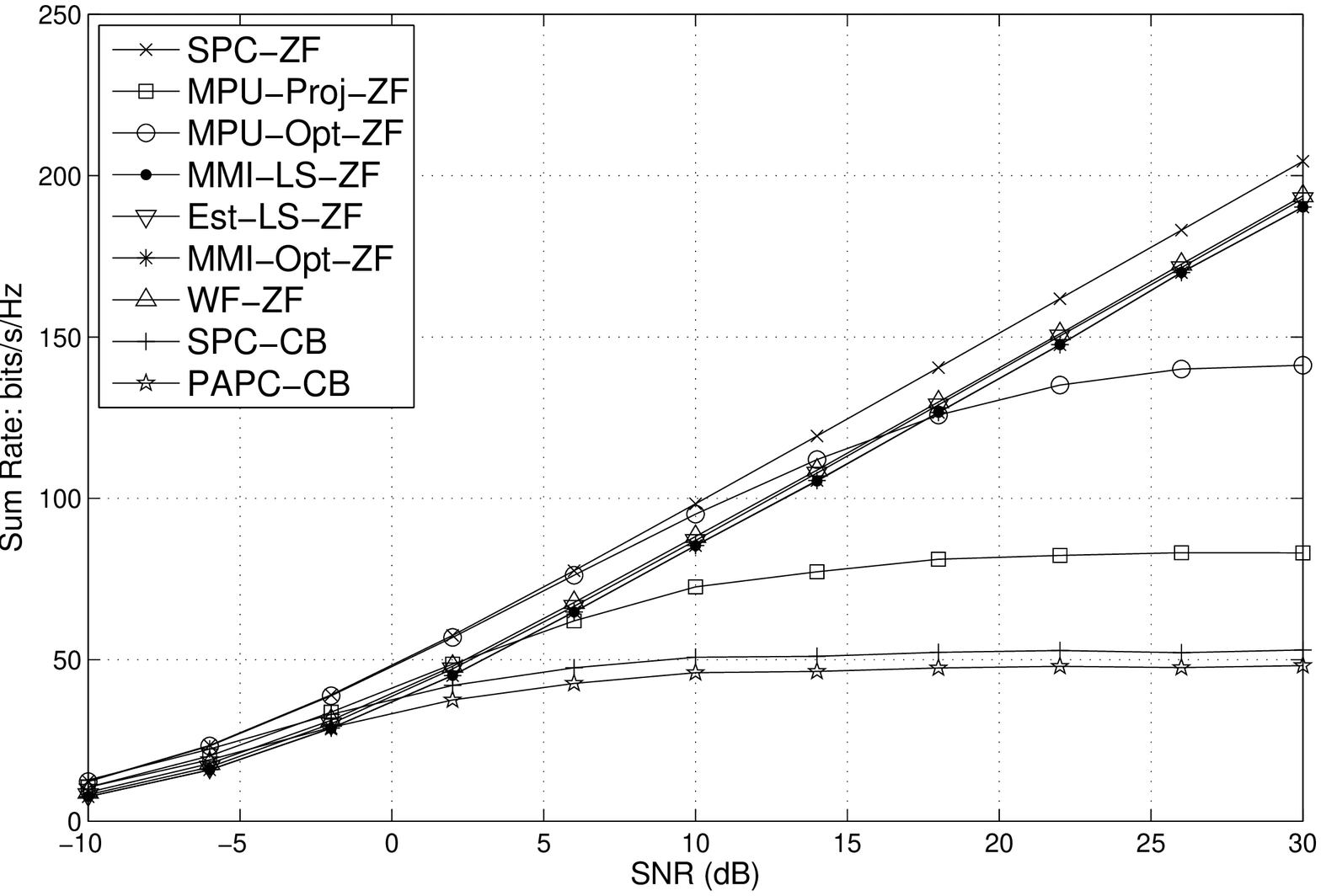}
\caption{Ideal CSI, sum rate vs. SNR, $K=16,M=128$.}
\end{figure}
\fi

\ifCLASSOPTIONonecolumn
\begin{figure}[!t]
\centering \includegraphics[width = 1.0\linewidth]{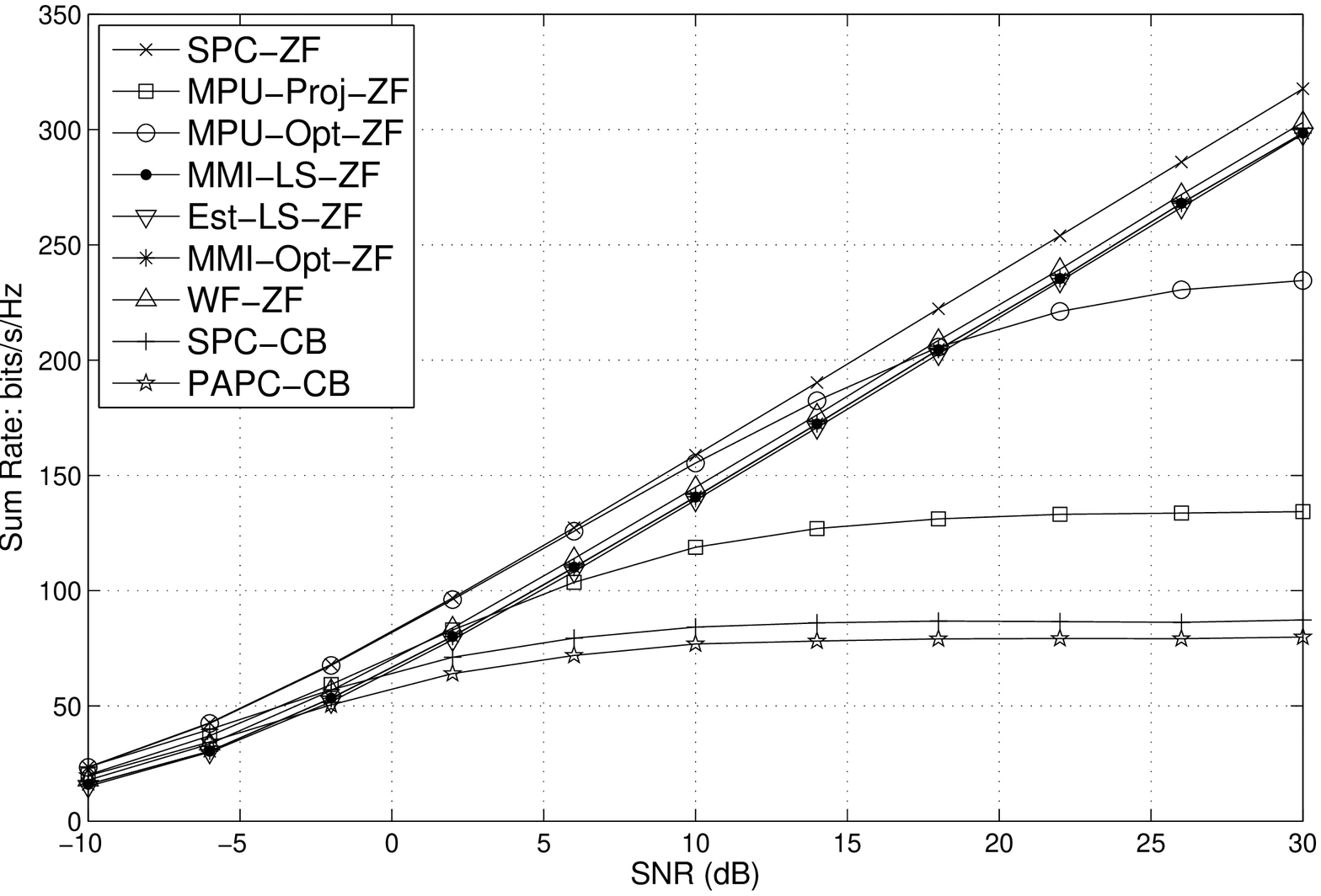}
\caption{Ideal CSI, sum rate vs. SNR, $K=24,M=256$.}
\end{figure}
\else
\begin{figure}[!t]
\centering \includegraphics[width = 1.0\linewidth]{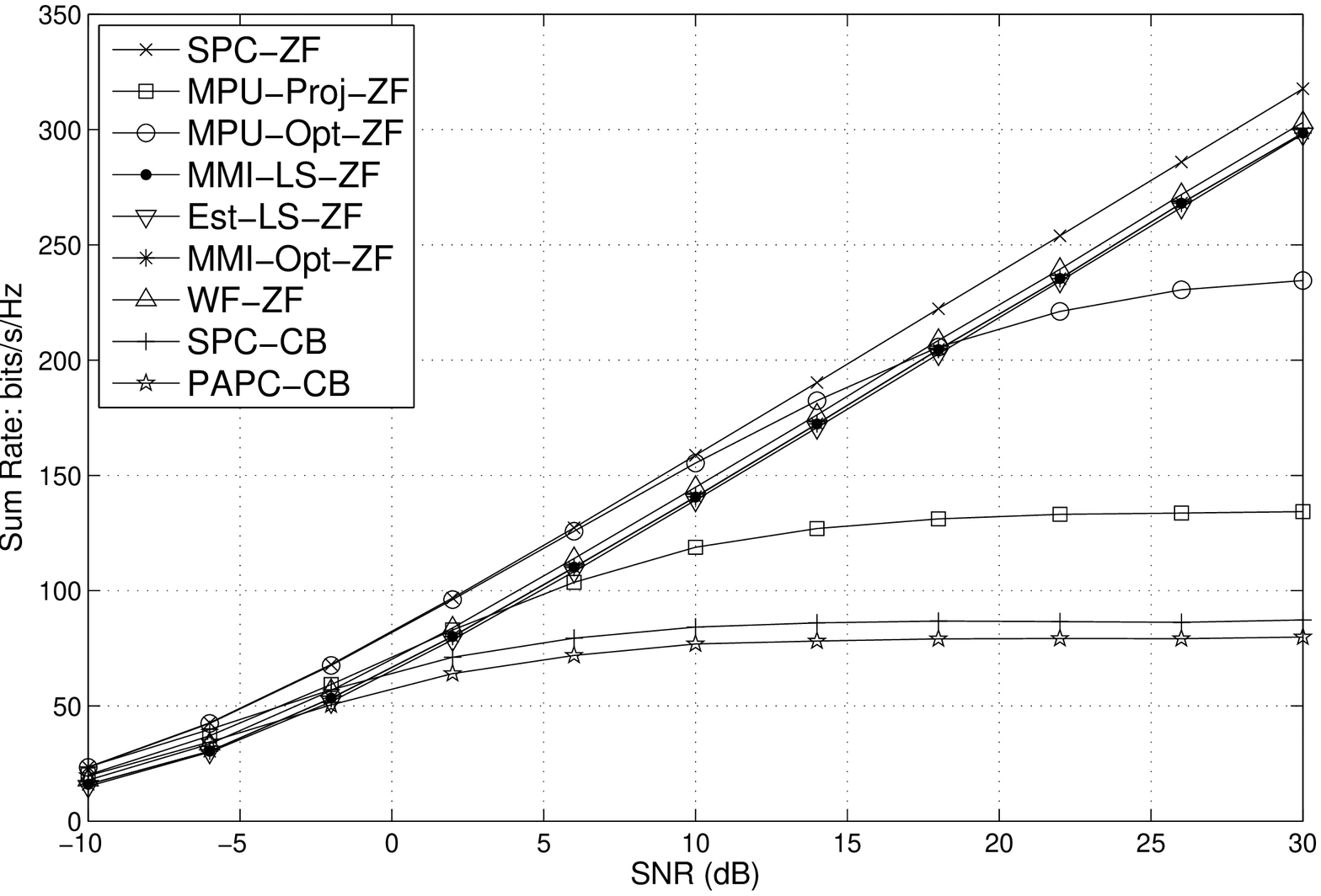}
\caption{Ideal CSI, sum rate vs. SNR, $K=24,M=256$.}
\end{figure}
\fi
In Fig. $5$ and Fig. $6$, the sum rate vs. SNR results are shown. Both figure shows that the two ZF-based methods maximizing the power utilization proposed in Section III achieve better performance compared to MMI-LS-ZF and MMI-Opt-ZF in the relatively low SNR region, i.e., less than around $2\mathrm{dB}$ and $15\mathrm{dB}$ for MPU-Proj-ZF and MPU-Opt-ZF respectively. In addition, MPU-Opt-ZF with the complexity of $O(m_{\mathrm{FNI}}MK)$ achieves significantly better performance than MPU-Proj-ZF with the complexity of $O(MK)$, which verifies the analysis in the subsection III-A. Moreover, MPU-Opt-ZF achieves the performance similar to SPC-ZF when the SNR is no more than around $10\mathrm{dB}$. As for the two ZF-based methods minimizing the multi-user interference proposed in Section IV, they achieve better performance compared to MPU-Proj-ZF and MPU-Opt-ZF in the relatively high SNR region, i.e., more than around $5\mathrm{dB}$ and $17\mathrm{dB}$ compared to MPU-Proj-ZF and MPU-Opt-ZF respectively. In addition, MMI-LS-ZF with the complexity of $O(MK)$ achieves the performance, which suffers an acceptable loss compared to SPC-ZF, similar to MMI-Opt-ZF with the complexity of $O[m_{\mathrm{NI}}(MK^2+K^3)]$ and WF-ZF with the complexity of $O[m_{\mathrm{WF}}(MK^2+K^3)]$. The reason is that, on the one hand, as the number of constraint equations is very large in (\ref{f18}), the optimal point of MMI-Opt-ZF is close to the point of MMI-LS-ZF. On the other hand, since the channel gains of different UEs approach to the same because of the law of large numbers in large-scale MIMO systems \cite{Marzetta_massive_MIMO_original,Hoydis_LPComp,Hoydis_massive_MIMO,Rusek_massive_MIMO_overview,Bjornson_arXiv,Larsson_massive_MIMO_overview}, WF-ZF offers almost the same solution as MMI-LS-ZF. Moreover, the approximation Est-LS-ZF achieves the performance close to MMI-LS-ZF, which verifies approximations (\ref{f24}) and (\ref{f29}). As for CB-PAPC proposed in Section VI, the performance loss compared to SPC-CB is acceptable, which verifies the around $0.35 \mathrm{bits/s/Hz}$ loss per user as discussed in Section VI.



\subsection{Non-ideal CSI}
In order to investigate the influence of non-ideal CSI, the sum rate vs. CSI error level results for $K=16$ and $M=128$  with the SNRs of $-10\mathrm{dB}$, $10\mathrm{dB}$, and $30\mathrm{dB}$ are shown in Fig. $7$, Fig. $8$ and Fig. $9$ respectively. In these figures, the parameter $\beta$ is used to denote the normalized correlation coefficient between ideal and measured CSI at a BS as in Fig. $3$ and Fig. $4$, where a larger value of $\beta$ results in a lower CSI error level. 

\ifCLASSOPTIONonecolumn
\begin{figure}[!t]
\centering \includegraphics[width = 1.0\linewidth]{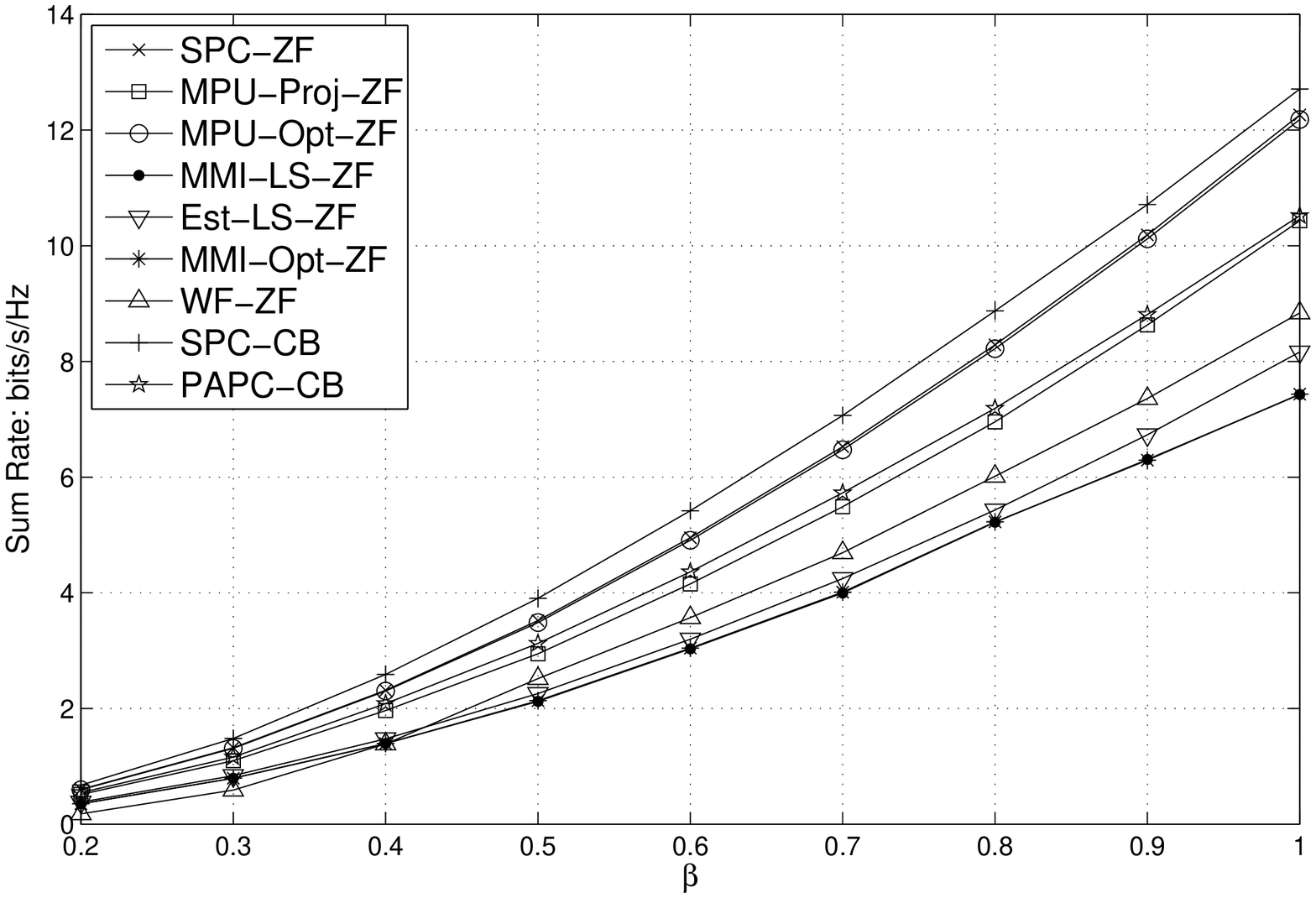}
\caption{ Sum rate vs. $\alpha$ when $\mathrm{SNR}=-10\mathrm{dB}$.}
\end{figure}
\else
\begin{figure}[!t]
\centering \includegraphics[width = 1.0\linewidth]{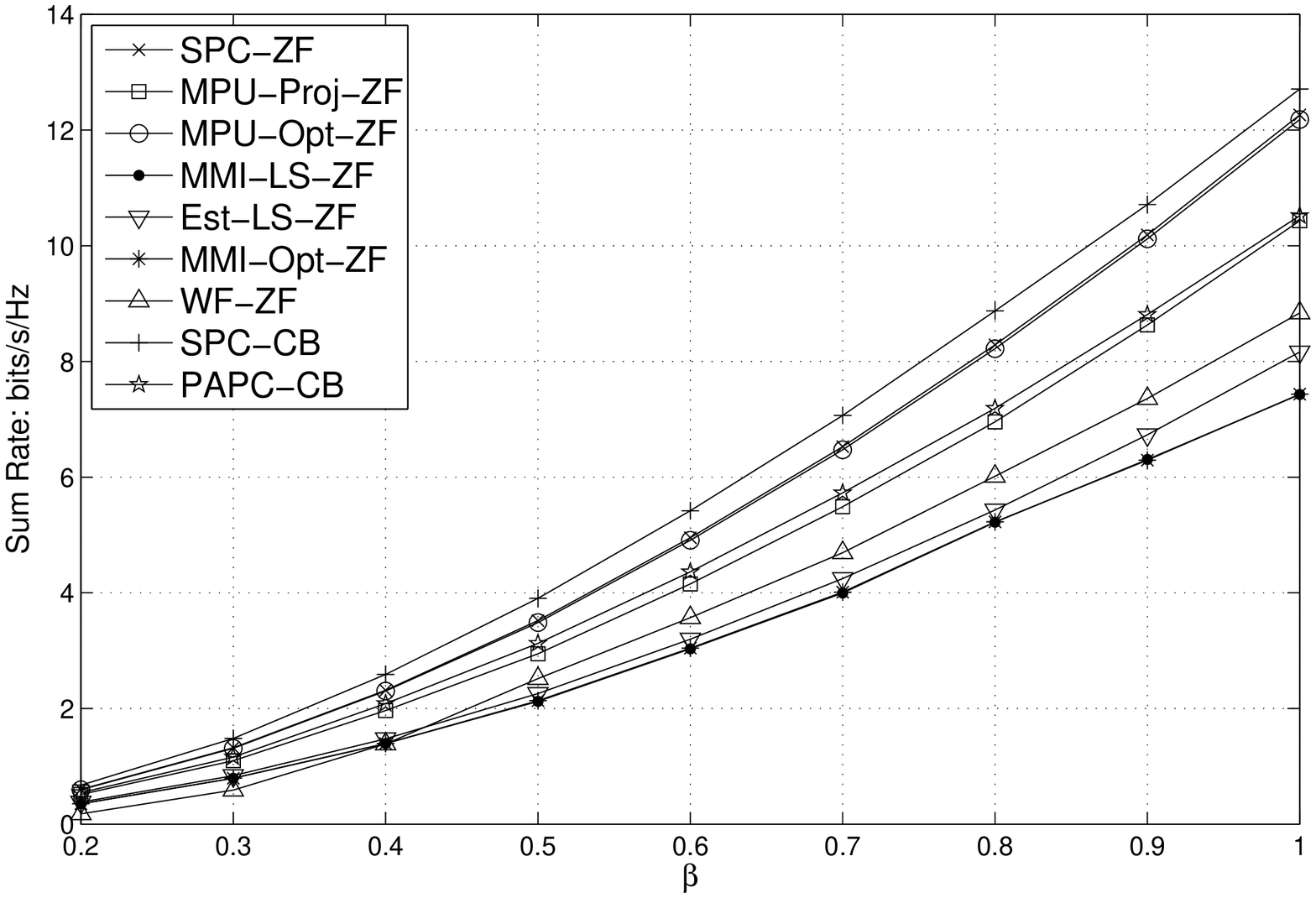}
\caption{ Sum rate vs. $\alpha$ when $\mathrm{SNR}=-10\mathrm{dB}$.}
\end{figure}
\fi

\ifCLASSOPTIONonecolumn
\begin{figure}[!t]
\centering \includegraphics[width = 1.0\linewidth]{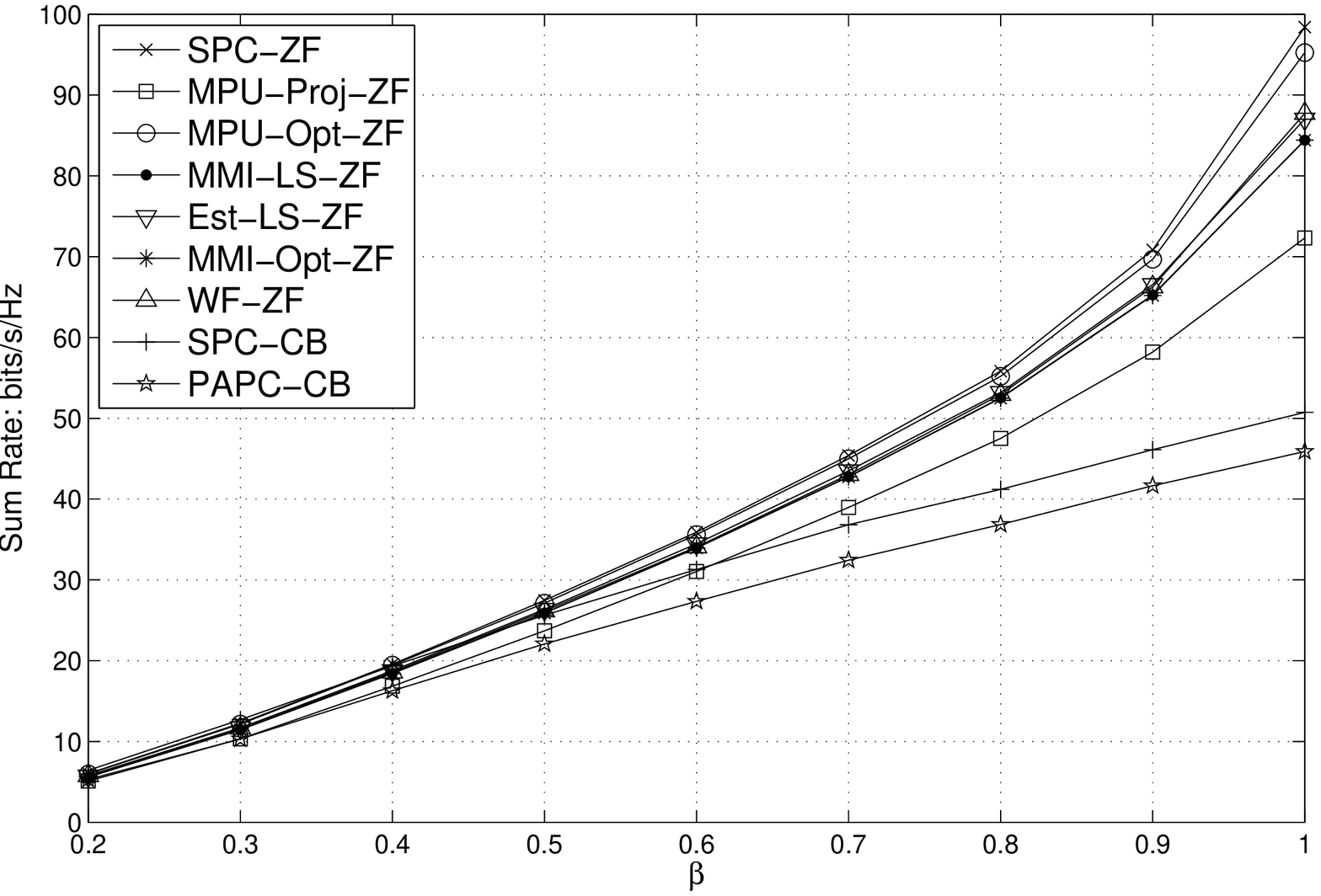}
\caption{Sum rate vs. $\alpha$ when $\mathrm{SNR}=10\mathrm{dB}$.}
\end{figure}
\else
\begin{figure}[!t]
\centering \includegraphics[width = 1.0\linewidth]{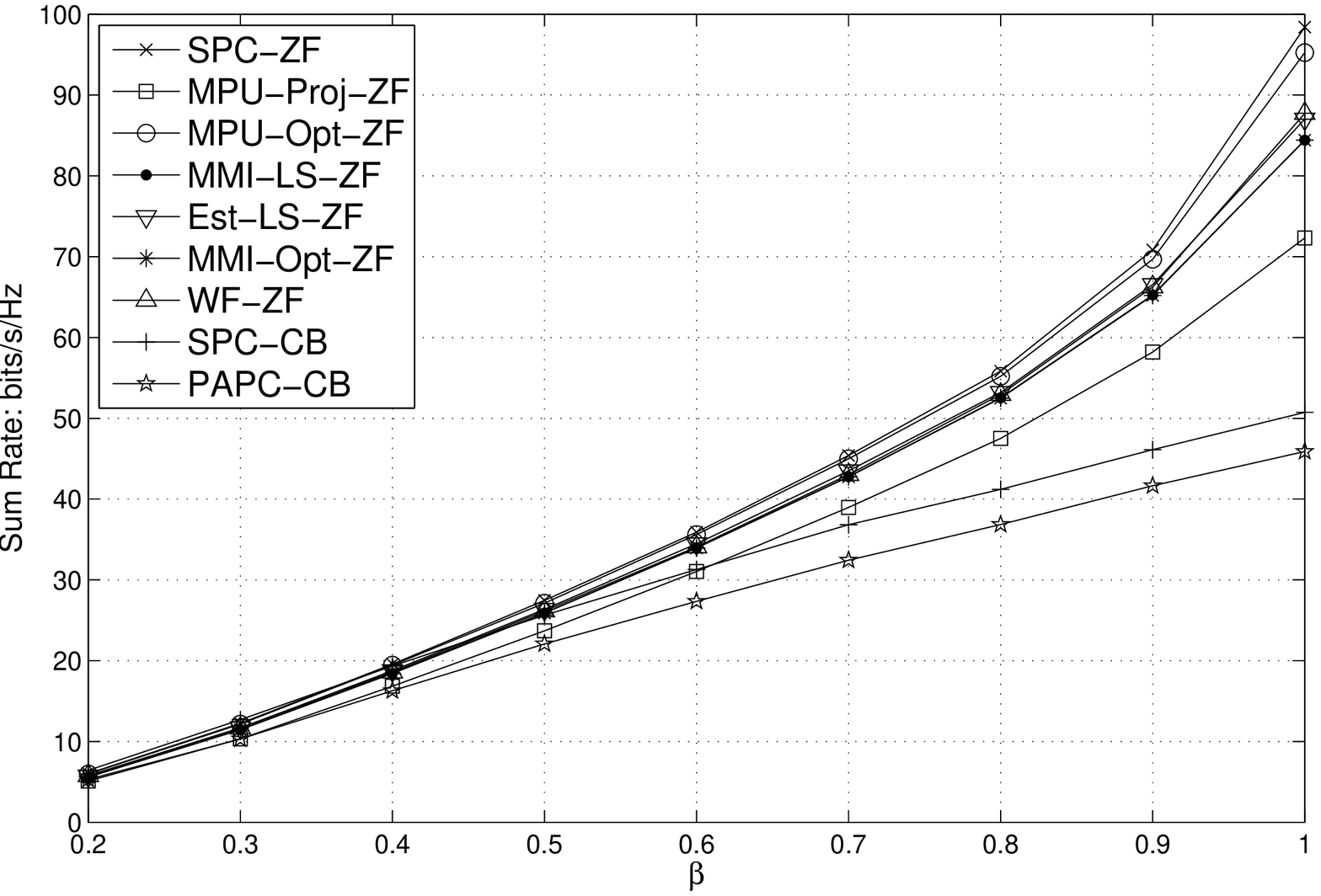}
\caption{Sum rate vs. $\alpha$ when $\mathrm{SNR}=10\mathrm{dB}$.}
\end{figure}
\fi

\ifCLASSOPTIONonecolumn
\begin{figure}[!t]
\centering \includegraphics[width = 1.0\linewidth]{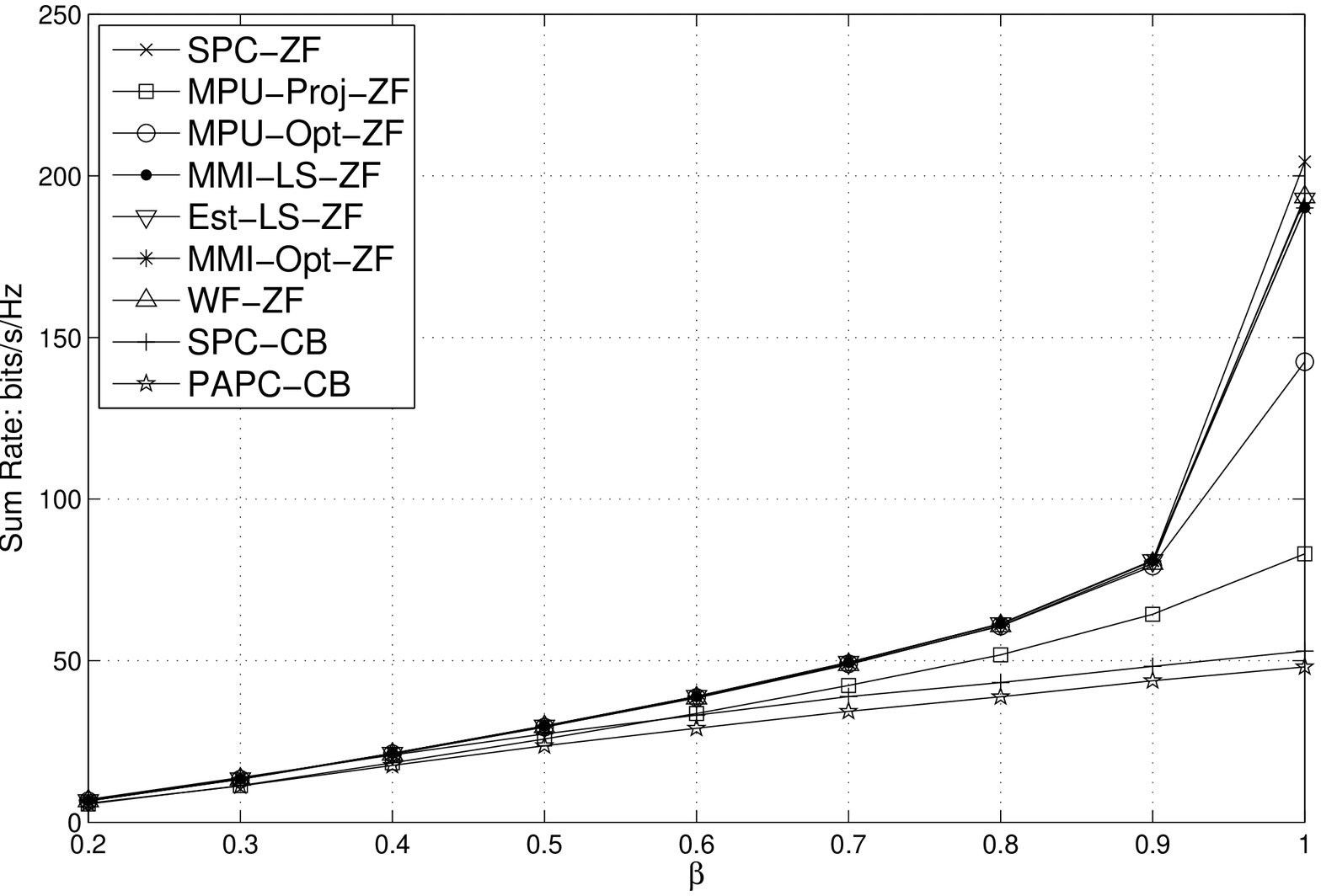}
\caption{Sum rate vs. $\alpha$ when $\mathrm{SNR}=30\mathrm{dB}$.}
\end{figure}
\else
\begin{figure}[!t]
\centering \includegraphics[width = 1.0\linewidth]{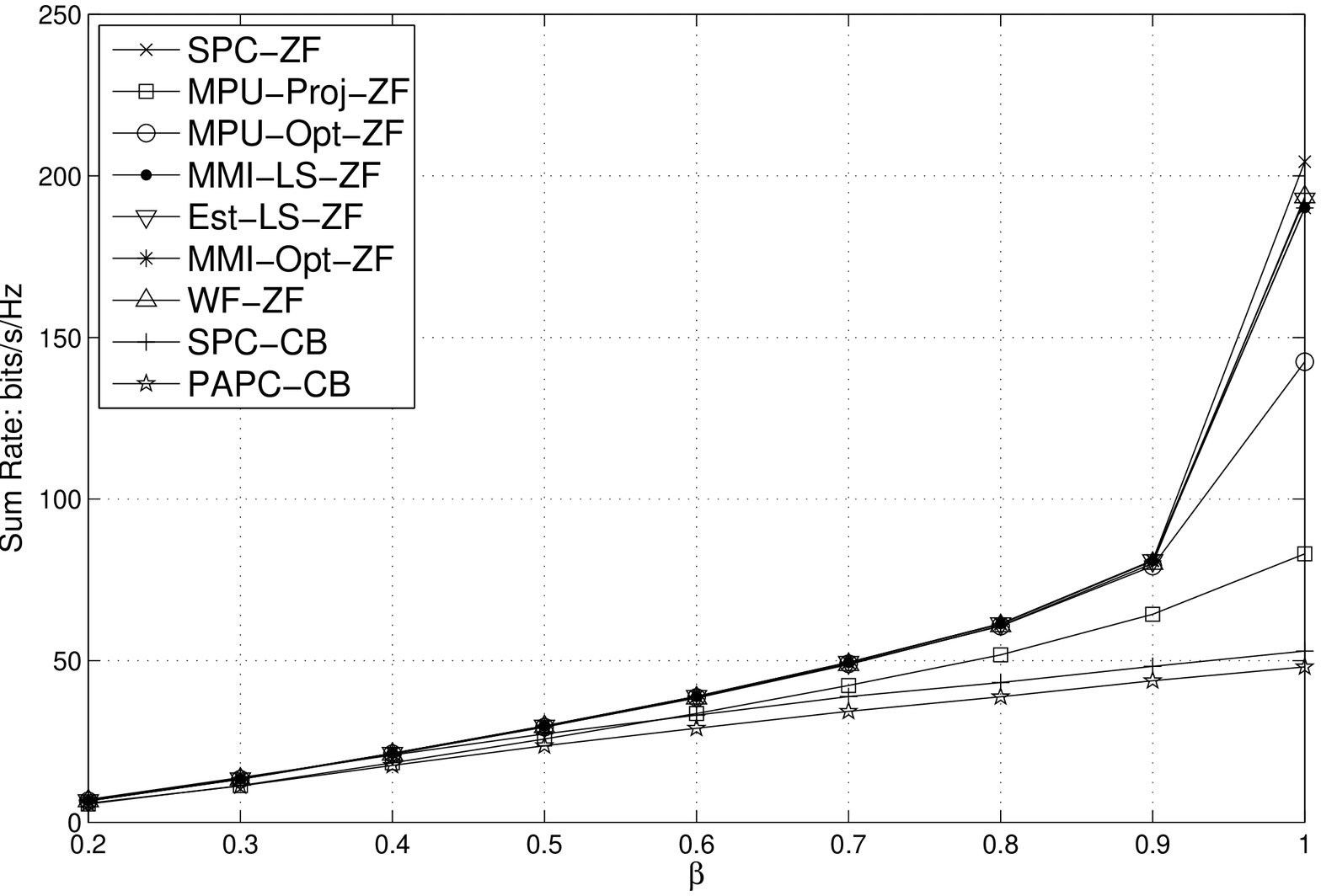}
\caption{Sum rate vs. $\alpha$ when $\mathrm{SNR}=30\mathrm{dB}$.}
\end{figure}
\fi

Fig. $7$ shows that when the is SNR $-10\mathrm{dB}$, for the four proposed ZF-based algorithms, MPU-Proj-ZF and MPU-Opt-ZF outperform MMI-LS-ZF and MMI-Opt-ZF, regardless of $\beta$. Especially, MPU-Opt-ZF achieves the best performance, which is almost the same as SPC-ZF, compared to MPU-Proj-ZF, MMI-LS-ZF, MMI-Opt-ZF, WF-ZF, and PAPC-CB. In addition, as the SNR is $-10\mathrm{dB}$, SPC-CB achieves better performance compared to SPC-ZF, which is consistent with the conventional understanding that CB offers higher performance than ZF in the relatively low SNR region \cite{Rusek_massive_MIMO_overview}. In the case of  PAPC, PAPC-CB achieves the performance better than MMI-LS-ZF, MMI-Opt-ZF, and WF-ZF, close to MPU-Proj-ZF, and only worse than MPU-Opt-ZF. Note that although PAPC-CB is worse than MPU-Opt-ZF, the complexity of CB is much smaller than ZF. Moreover, the approximation Est-LS-ZF is still very close to MMI-LS-ZF. 

Similarly. Fig. $8$ shows that when the SNR is $10\mathrm{dB}$, MPU-Opt-ZF still achieves the best performance, which is almost the same as SPC-ZF, compared to MPU-Proj-ZF, MMI-LS-ZF, MMI-Opt-ZF, WF-ZF, and PAPC-CB. In addition, MMI-LS-ZF achieves the performance similar to MMI-Opt-ZF and WF-ZF regardless of $\beta$ when SNR is $10\mathrm{dB}$, which suffers only a small loss compared to MPU-Opt-ZF and is significantly better than MPU-Proj-ZF and PAPC-CB. Note that the approximation Est-LS-ZF is almost the same as MMI-LS-ZF in this case. Moreover, PAPC-CB with an acceptable loss compared to SPC-CB achieves the worst performance regardless of $\beta$. The reason is that the inter-user interference instead of the power of the target signal becomes the dominate factor of the post-receiving SINR when SNR increases to $10\mathrm{dB}$.  

Finally, Fig. $9$ shows that when the SNR is $30\mathrm{dB}$, MMI-LS-ZF still achieves the performance close to MMI-Opt-ZF and WF-ZF and better than MPU-Proj-ZF and PAPC-CB. In addition, when $\beta<0.9$, MMI-LS-ZF achieves almost the same performance as SPC-ZF and MPU-Opt-ZF. When $0.9 \leq \beta \leq 1$, MMI-LS-ZF suffers only a very small loss compared to SPC-ZF and is significantly better than MPU-Opt-ZF, which indicates that MMI-LS-ZF is slightly worse than SPC-ZF and significantly better than MPU-Opt-ZF only when the CSI is relatively accurate. Note that the approximation Est-LS-ZF is almost the same as MMI-LS-ZF in this case. Moreover, PAPC-CB with an acceptable loss compared to SPC-CB achieves the worst performance regardless of $\beta$. 

In summary, when $\beta<0.9$, under the PAPC, MPU-Opt-ZF always achieves the best performance regardless of SNR. In the relatively high SNR region, MMI-LS-ZF, MMI-Opt-ZF, and WF-ZF achieve the performance similar to MPU-Opt-ZF, but MPU-Opt-ZF outperforms them in the relatively low SNR region. When $0.9 < \beta\leq 1$, the results are similar to the ideal CS case. Specifically, MPU-Opt-ZF achieves the best performance when the SNR is relatively small
. Otherwise, MMI-LS-ZF, MMI-Opt-ZF, and WF-ZF achieve the best performance, where MMI-LS-ZF has the lowest complexity. Note that the best choices for difference cases all achieve acceptable or negligible losses compared to SPC-ZF, which verifies the effectiveness of MPU-Opt-ZF, MMI-LS-ZF, and MMI-Opt-ZF. Although MPU-Proj-ZF is not the best choice in various values of SNR and $\beta$, its solution offers an initial value to MPU-Opt-ZF to ensure its rapid convergence. In addition, the error of the approximation Est-LS-ZF to MMI-LS-ZF 
is negligible, which 
verifies approximations (\ref{f24}) and (\ref{f29}). Moreover, the performance loss of PAPC-CB compared to SPC-CB is acceptable, which verifies the effectiveness of PAPC-CB. 

\subsection{Application in Practical Systems}
In practical systems, unavoidable CSI errors exist, which is a more serious issue for large-scale MIMO systems. On the one hand, highly accurate CSI measurement means huge pilot overhead. On the other hand, the inevitable calibration error of the Time Domain Duplexing (TDD) mode, which is the mainly considered operation mode for large-scale MIMO systems \cite{Marzetta_massive_MIMO_original,Hoydis_LPComp,Hoydis_massive_MIMO,Rusek_massive_MIMO_overview,Bjornson_arXiv,Larsson_massive_MIMO_overview}, worsens the CSI in the downlink transmission. Therefore, an adaptive power allocation method is the best choice for practical large-scale MIMO systems. When the CSI error is relatively large, i.e., the post-receiving SINR of uplink pilot signal is relatively low such that $\beta<0.9$, MPU-Opt-ZF is employed regardless of the downlink SNR. Otherwise, MMI-LS-ZF is employed for its simplicity when the downlink SNR is relatively large
, while MPU-Opt-ZF is employed 
when the downlink SNR is relatively low
. Note that ZF involves large amount of computation complexity to obtain the inverse matrix, which could cause large processing delay at the BS. Hence, CB could be a more reasonable choice for UEs that move fast for its simplicity. In this case, CB-PAPC is applied. 
   
\section{Conclusions}
In this paper, we provide four ZF-based power allocation methods according to two different criteria and one CB-based method to solve the power allocation problem under the PAPC for large-scale MIMO systems. In addition, the sum rate loss of ZF-based linear scaling relative to the virtually optimal case of ZF under the SPC, as well as the sum rate loss of the CB-based method relative to CB under the SPC, are derived. Simulation results show that with relatively accurate CSI, i.e., $\beta \geq 0.9$, the feasible Newton iterative method based on maximum power utilization leads to the largest average achievable sum rate in the relatively low SNR region, which is almost the same as the virtually optimal case of ZF under the SPC. In the relatively high SNR region with relatively accurate CSI, although the two proposed methods based on minimum multi-user both achieves similar best achievable sum rates with only small losses compared to
ZF under the SPC, along with the water-filling based method employed in \cite{Boccardi_ZF_PAPC,Ohwatari_PAPC}, the linear scaling method has the lowest complexity. If the CSI is relatively inaccurate, i.e., $\beta<0.9$, 
the feasible Newton iterative method is the best choice with almost the same
performance as ZF under the SPC regardless of the SNR. 
In addition, the maximum SINR gap between linear scaling and the optimal case of ZF under the SPC is about $f_{\mathrm{LS}}(K,M)$ for ideal CSI, and it decreases to $0$ as the CSI error increases. Furthermore, the proposed CB-based power allocation 
suffers little throughput loss relative to the 
case of CB under the SPC. The results could provide useful references for practice.

\bibliographystyle{IEEEtran}
\bibliography{IEEEabrv,Mybib}

\begin{thebibliography}{10}
\providecommand{\url}[1]{#1}
\csname url@samestyle\endcsname
\providecommand{\newblock}{\relax}
\providecommand{\bibinfo}[2]{#2}
\providecommand{\BIBentrySTDinterwordspacing}{\spaceskip=0pt\relax}
\providecommand{\BIBentryALTinterwordstretchfactor}{4}
\providecommand{\BIBentryALTinterwordspacing}{\spaceskip=\fontdimen2\font plus
\BIBentryALTinterwordstretchfactor\fontdimen3\font minus
  \fontdimen4\font\relax}
\providecommand{\BIBforeignlanguage}[2]{{%
\expandafter\ifx\csname l@#1\endcsname\relax
\typeout{** WARNING: IEEEtran.bst: No hyphenation pattern has been}%
\typeout{** loaded for the language `#1'. Using the pattern for}%
\typeout{** the default language instead.}%
\else
\language=\csname l@#1\endcsname
\fi
#2}}
\providecommand{\BIBdecl}{\relax}
\BIBdecl

\bibitem{Marzetta_massive_MIMO_original}
T.~L. Marzetta, ``{Noncooperative Cellular Wireless with Unlimited Numbers of
  Base Station Antennas},'' \emph{{IEEE} Trans. Wireless Commun.}, vol.~9,
  no.~11, pp. 3590--3600, Nov. 2010.

\bibitem{Hoydis_LPComp}
J.~Hoydis, S.~Brink, and M.~Debbah, ``{Comparison of Linear Precoding Schemes
  for Downlink Massive MIMO},'' in \emph{{Proc. IEEE ICC 2012}}, Ottawa,
  Canada, Jun. 2012, pp. 2135--2139.

\bibitem{Hoydis_massive_MIMO}
------, ``{Massive MIMO in the UL/DL of Cellular Networks: How Many Antennas Do
  We Need?}'' \emph{IEEE Sel. Areas Commun.}, vol.~31, no.~2, pp. 160--171,
  Feb. 2013.

\bibitem{Rusek_massive_MIMO_overview}
F.~Rusek, D.~Persson, B.~K. Lau, E.~G. Larsson, T.~L. Marzetta, O.~Edfors, and
  F.~Tufvesson, ``{Scaling up MIMO: Opportunities and Challenges with Very
  Large Arrays},'' \emph{{IEEE} Signal Process. Mag.}, vol.~30, no.~1, pp.
  40--46, Jan. 2013.

\bibitem{Bjornson_arXiv}
\BIBentryALTinterwordspacing
E.~Bjornson, L.~Sanguinetti, J.~Hoydis, and M.~Debbah. (2013) {Design
  Multi-User MIMO for energy Efficiency: When is Massive MIMO the Answer?}
  arXiv:1310.3843. [Online]. Available: \url{http://arxiv.org}
\BIBentrySTDinterwordspacing

\bibitem{Larsson_massive_MIMO_overview}
E.~G. Larsson, F.~Tufvesson, O.~Edfors, and T.~L. Marzetta, ``{Massive MIMO for
  Next Generation Wireless Systems},'' \emph{{IEEE} Commun. Mag.}, vol.~52,
  no.~2, pp. 186--195, Feb. 2014.

\bibitem{Yu_PAPC}
W.~Yu and T.~Lan, ``{Transmitter Optimization for the Multi-Antenna Downlink
  With Per-Antenna Power Constraints},'' \emph{{IEEE} Trans. Signal Process.},
  vol.~55, no.~6, pp. 2646--2660, Jun. 2007.

\bibitem{Costa_DPC}
M.~H.~M. Costa, ``{Writing on Dirty Paper},'' \emph{{IEEE} Trans. Inf. Theory},
  vol.~29, no.~3, pp. 439--441, May 1983.

\bibitem{Caire_MIMO_Broadcast}
G.~Caire and S.~Shamai, ``{On the Achievable Throughput of A Multiantenna
  Gaussian Broadcast Channel},'' \emph{{IEEE} Trans. Inf. Theory}, vol.~49,
  no.~7, pp. 1691--1706, Jul. 2003.

\bibitem{Viswanath_Broadcast}
P.~Viswanath and D.~N.~C. Tse, ``{Sum Capacity of the Vector Gaussian Broadcast
  Channel and Uplink-Downlink Duality},'' \emph{{IEEE} Trans. Inf. Theory},
  vol.~49, no.~8, pp. 1912--1921, Aug. 2003.

\bibitem{Yu_Broadcast}
W.~Yu and J.~M. Cioffi, ``{Sum Capacity of Gaussian Vector Broadcast
  Channels},'' \emph{{IEEE} Trans. Inf. Theory}, vol.~50, no.~9, pp.
  1875--1892, Sep. 2004.

\bibitem{Lee_DPC_LP}
J.~Lee and N.~Jindal, ``{High SNR Analysis for MIMO Broadcast Channels: Dirty
  Paper Coding Versus Linear Precoding},'' \emph{{IEEE} Trans. Inf. Theory},
  vol.~53, no.~12, pp. 4787--4792, Dec. 2007.

\bibitem{Christopoulos_PAPC}
\BIBentryALTinterwordspacing
D.~Christopoulos, S.~Chatzinotas, and B.~Ottersten. (2014) {Sum Rate Maximizing
  Multigroup Multicast Beamforming under Per-antenna Power Constraints}.
  arXiv:1407.0005. [Online]. Available: \url{http://arxiv.org}
\BIBentrySTDinterwordspacing

\bibitem{Wiesel_ZFP}
A.~Wiesel, Y.~C. Eldar, and S.~Shamai, ``{Zero-Forcing Precoding and
  Generalized Inverses},'' \emph{{IEEE} Trans. Signal Process.}, vol.~56,
  no.~9, pp. 4409--4418, Sep. 2008.

\bibitem{Shepard_Argos}
C.~Shepard, H.~Yu, N.~Anand, L.~E. Li, T.~Marzetta, R.~Yang, and L.~Zhong,
  ``{Argos: Practical Many-Antenna Base Stations},'' in \emph{{Proc. MobiCom
  12}}, Istanbul, Turkey, Aug. 2012.

\bibitem{Marzetta_ZF_CB}
H.~Yang and T.~L. Marzetta, ``{Performance of Conjugate and Zero-Forcing
  Beamforming in Large-scale Antenna Systems},'' \emph{{IEEE} J. Sel. Areas
  Commun.}, vol.~31, no.~2, pp. 172--179, Feb. 2013.

\bibitem{Vandenberghe_MAXDET}
L.~Vandenberghe, S.~Boyd, and S.-P. Wu, ``{Determinant Maximization with Linear
  Matrix Inequality Constraints},'' \emph{{SIAM Journal on Matrix Analysis and
  Applications}}, vol.~19, no.~2, p. 499–533, Apr. 1998.

\bibitem{Boccardi_ZF_PAPC}
F.~Boccardi and H.~Huang, ``{Zero-Forcing Precoding for the MIMO Broadcast
  Channel under Per-Antenna Power Constraints},'' in \emph{{Proc. IEEE SPAWC
  2006}}, Cannes, France, Jul. 2006.

\bibitem{Ohwatari_PAPC}
Y.~Ohwatari, A.~Benjebbour, J.~Hagiwara, and T.~Ohya, ``{Reduced-Complexity
  Transmit Power Optimization Techniques for Multiuser MIMO with Per-Antenna
  Power Constraint},'' in \emph{{Proc. Allerton 08}}, Urbana-Champaign, IL,
  USA, Sep. 2008, pp. 34--38.

\bibitem{Horn_Matrix_Analysis}
R.~A. Horn and C.~R. Johnson, \emph{{Matrix Analysis}}.\hskip 1em plus 0.5em
  minus 0.4em\relax {Cambridge Univeristy Press}, 1990, ch.~6.

\bibitem{NeumannSeries}
N.~Suzuki, ``{On the convergence of Neumann series in Banach space},''
  \emph{{Mathematische Annalen}}, vol. 220, no.~2, pp. 143--146, 1976.

\bibitem{Boyd_CO}
S.~Boyd and L.~Vandenberghe, \emph{{Convex Optimization}}.\hskip 1em plus 0.5em
  minus 0.4em\relax {Cambridge Univeristy Press}, 2004.

\bibitem{Tulino_RMT}
{A. M. Tulino and S. Verd{\'u}}, ``{Random Matrix Theory and Wireless
  Communications},'' \emph{Foundations and Trends® in Communications and
  Information Theory}, vol.~1, no.~1, p. 1‐182, Jun. 2004.

\bibitem{Guest_CF}
P.~G. Guest, \emph{{Numerical Methods of Curve Fitting}}.\hskip 1em plus 0.5em
  minus 0.4em\relax {Cambridge Univeristy Press}, 2012.

\bibitem{Zhu_Grassmann_arXiv}
\BIBentryALTinterwordspacing
D.~Zhu, B.~Li, and P.~Liang. (2014) {Normalized Volume of Hyperball in Complex
  Grassmann Manifold and Its Application in Large-Scale MU-MIMO Communication
  Systems}. arXiv:1402.4543. [Online]. Available: \url{http://arxiv.org}
\BIBentrySTDinterwordspacing

\bibitem{Matrix_Computation}
G.~H. Golub and C.~F.~V. Loan, \emph{{Matrix Computation}}, 3rd~ed.\hskip 1em
  plus 0.5em minus 0.4em\relax {Johns Hopkins University Press}, 1996, ch.~9.

\bibitem{Proakis_DC}
J.~Proakis and M.~Salehi, \emph{{Digital Communications}}, 5th~ed.\hskip 1em
  plus 0.5em minus 0.4em\relax {McGraw-Hill}, 2008.

\bibitem{Madhow_ComFund}
U.~Madhow, \emph{{Fundamentals of Digital Communication}}.\hskip 1em plus 0.5em
  minus 0.4em\relax {Cambridge Univeristy Press}, 2008, ch.~6.

\end{thebibliography}
\end{document}